# Micromechanics of dislocations in solids: $J$-, $M$-, and $L$-integrals and their fundamental relations

Eleni Agiasofitou, Markus Lazar[*]

*Heisenberg Research Group, Department of Physics, Darmstadt University of Technology, Hochschulstr. 6, D-64289 Darmstadt, Germany*

## Abstract

The aim of the present work is the unification of incompatible elasticity theory of dislocations and Eshelbian mechanics leading naturally to Eshelbian dislocation mechanics. In such a unified framework, we explore the utility of the $J$-, $M$-, and $L$-integrals. We give the physical interpretation of the $M$-, and $L$-integrals for dislocations, connecting them with established quantities in dislocation theory such as the interaction energy and the $J$-integral of dislocations, which is equivalent to the well-known Peach-Koehler force. The $J$-, $M$-, and $L$-integrals for dislocations have been studied in the framework of three-dimensional, incompatible, linear elasticity. First of all, the general formulas of the $J$-, $M$-, and $L$-integrals for dislocations are given. Next, the examined integrals are specified for straight dislocations. Finally, the explicit formulas of the $J$-, $M$-, and $L$-integrals are calculated for straight (screw and edge) dislocations in isotropic materials. The obtained results reveal the physical interpretation and significance of the $M$-, and $L$-integrals for straight dislocations. The $M$-integral between two straight dislocations (per unit dislocation length) is equal to the half of the interaction energy of the two dislocations (per unit dislocation length) depending on the distance and on the angle, plus twice the corresponding pre-logarithmic energy factor. The $L_3$-integral between two straight dislocations is the $z$-component of the configurational vector moment or the rotational moment about the $z$-axis caused by the interaction between the two dislocations. Fundamental relations between the $J$-, $M$-, and $L_3$-integrals are derived showing the inherent connection between them. The relations connecting directly the $J$-, $M$-, and $L_3$-integrals with the interaction energy are obtained. These relations have been proven to be of great significance. Since based on them; the interpretation of the $J$-, and $L_3$-integrals as translational and rotational energy-release, respectively, is achieved, and secondly a stability criterion for straight edge dislocations is formulated in terms of the $J_k$-integral, revealing the physical importance of the considered integrals.

## 1. Introduction

Dislocations are one of the most important defects in solids, since they influence the properties of a crystal; not only the mechanical, but also the electric, magnetic, optic, and semi-conducting properties as well as the growth of a crystal (e.g., Nabarro (1967)). Dislocations cause plasticity and hardening in crystals (e.g., Cottrell (1953); Lardner (1974)). Consequently, the physics of the interaction between dislocations is an important research field and for that reason this work is focused on the study of the $J$-, $M$-, and $L$-integrals between two dislocations. The interaction force between dislocations is the so-called Peach-Koehler

---

[*]Corresponding author.
*Email addresses:* agiasofitou@mechanik.tu-darmstadt.de (Eleni Agiasofitou), lazar@fkp.tu-darmstadt.de (Markus Lazar)

force (Peach and Koehler, 1950). Rogula (1977) and Kirchner (1984, 1999) gave a field-theoretical derivation of the Peach-Koehler force as the divergence of the so-called Eshelby stress tensor (Eshelby, 1975), using the framework of dislocation theory including incompatible distortion tensors. Rice (1985); Weertman (1996) and Kirchner (1999) found that the Peach-Koehler force is equivalent to the $J$-integral of dislocations. In general, the Peach-Koehler force is a configurational or material force. Material forces are forces acting on fields in the sense of Lorentzian forces, for example the static Lorentz force is the interaction force on an electric charge in the presence of an electric field. Material forces should not be confused with forces in the sense of Newtonian forces, acting on masses in the second Newton law; where a force is defined as mass times acceleration (see also Kröner (1993); Maugin (1993)). In elasticity, Newtonian forces appear as the divergence of the Cauchy stress tensor, whereas material forces appear as the divergence of the Eshelby stress tensor (static energy-momentum tensor). Material forces are forces that do act on defects such as dislocations, disclinations, point defects, cracks, etc. The physical meaning of the Peach-Koehler force as the interaction force acting on a dislocation, defined by the dislocation density tensor due to the stress field caused by another dislocation (or another defect) in an elastic material, is well-established in classical dislocation theory, and can be found in many textbooks on dislocations (e.g., Landau and Lifschitz (1986); Lardner (1974); Hirth and Lothe (1982); Li and Wang (2008)). In the framework of the dislocation gauge field theory, Agiasofitou and Lazar (2010) proved that the Peach-Koehler force is, in particular, the interaction force between the elastic and the dislocation subsystems enlightening that the Peach-Koehler force is self-equilibrating. Moreover, they have proven that the Peach-Koehler force is a material force caused by dislocations as incompatibilities.

The $J$-integral became originally famous in fracture mechanics. Rice (1968) used the Eshelby stress tensor to derive a certain conservation law which is known as the $J$-integral. Seven conservation laws of elasticity, which are related to translational, rotational, and scaling symmetries, were originally derived by Günther (1962), and Knowles and Sternberg (1972). The corresponding conservation integrals are the $J$-, $L$-, and $M$-integrals introduced by Budiansky and Rice (1973). Budiansky and Rice (1973) were the first to give a physical interpretation to the $J$-, $L$-, and $M$-integrals as the energy-release rates per unit cavity translation, rotation, and expansion, respectively. An important contribution to connect the $J$-, $L$-, and $M$-integrals with the energy-release rates for cracks and the relations between them, was given in a series of papers by Golebiewska-Herrmann and Herrmann (1981b); Pak *et al.* (1983), and Eischen and Herrmann (1987). Golebiewska-Herrmann and Herrmann (1981b) showed that the $L$-integral represents the rotational energy-release (rate)[1] induced from the rotation of a central plane crack. Chen (2002) (see also Chen and Lu (2003)) was the first to show that the $M$-integral equals twice the change of the total potential energy owed to single cracking of a central crack in a plane elastic body. This conclusion was implied by Budiansky and Rice (1973) and Golebiewska-Herrmann and Herrmann (1981b), but written explicitly by Chen (2002) emphasizing for the first time this physical interpretation of the $M$-integral. An interesting overview about conservation laws, $J$-, $L$-, and $M$-integrals as well as related concepts with various applications to defect mechanics in the framework of compatible linear elasticity is given by Kienzler and Herrmann (2000) and Chen (2002).

Using elasticity as field theory, Lazar and Kirchner (2007a) derived the $J$-, $L$-, and $M$-integrals for gradient elasticity theory of non-homogeneous, incompatible, linear, anisotropic media. For the first time deriving the general expressions of the $L$-, and $M$-integrals for dislocations, for both elasticity theory and gradient elasticity theory. For compatible micropolar elasticity, the $J$-, $L$-, and $M$-integrals were given by Lubarda and Markenscoff (2003). For incompatible micropolar elasticity including dislocations and disclinations, the $J$-, $L$-, and $M$-integrals were derived by Lazar and Kirchner (2007b). For incompatible micromorphic elasticity with dislocations, disclinations, and point defects the $J$-integral including the Peach-Koehler force as well as the Mathisson-Papapetrou force were given by Lazar and Maugin (2007). As mentioned above, the $J$-integral of dislocations in incompatible elasticity is equivalent to the Peach-Koehler force, which appears as the divergence of the Eshelby stress tensor. In the framework of configurational or

---

[1] Attention should be payed to the terminology "energy-release rate", since in the literature the expression "rate" is often used abusively in statics. For reasons of clarity, the terminology "energy-release" is used in this work, since it is in the framework of elastostatics; letting the terminology "energy-release rate" available for elastodynamics.

Eshelbian mechanics, Lazar and Kirchner (2007a) showed that for dislocations the $\boldsymbol{L}$-integral is equivalent to configurational or material vector moment, which appears as the divergence of the angular momentum tensor and the $M$-integral is equivalent to configurational or material work, which appears as the divergence of the scaling flux vector of dislocations.

The question that arises now is what is the specific physical meaning of the $M$-, and $\boldsymbol{L}$-integrals for dislocations. The first single work posing the question "What is $M$ for a dislocation?" goes back to Rice in 1985. Rice (1985) studied the $M$-integral when centered on a dislocation line in the framework of two-dimensional, compatible, linear elasticity and found that the $M$-integral is equal to the "dislocation energy factor" (see also Asaro and Lubarda (2006)). However, a dislocation is a line defect in a three-dimensional crystal and the expression of the $M$-integral is different in two- and three-dimensions, since the $M$-integral depends on the space dimension. In this work, we calculate explicitly the $M$-integral of two dislocations in the framework of three-dimensional, incompatible, linear elasticity revealing its real physical meaning. We show for the first time in dislocation theory that the $M$-integral between two straight dislocations (per unit dislocation length) is equal to the half of the interaction energy of the two dislocations (per unit dislocation length), which depends on the distance and on the angle, plus twice the corresponding pre-logarithmic energy factor[2]. Therefore, the $M$-integral of two straight dislocations has the physical interpretation of the interaction energy between the two dislocations depending on the distance and on the angle. This physical interpretation is in full agreement with the physical interpretation of the $M$-integral in fracture mechanics as we already discussed above that it was pointed out by Chen (2002).

The $\boldsymbol{L}$-integral for dislocations is more or less unknown in the scientific literature. Here, we prove that the $L_3$-integral has the physical interpretation of a torque or rotational moment about the $z$-axis between the two dislocations. Moreover, for the first time in the literature of dislocations, the $L_3$-integral is directly related with the rotational energy-release for straight dislocations. The rotational energy-release (rate) has been defined up to now only for cracks (Golebiewska-Herrmann and Herrmann, 1981b; Agiasofitou and Kalpakides, 2006). The interesting feature arising from the calculations is that the $L_3$-integral between two straight dislocations is a function of $\sin 2\varphi$ like the $L_3$-integral for cracks and cavities under various loading conditions as given by Pak *et al.* (2012).

Furthermore, to the best of our knowledge, this is the first work in the literature of dislocations where fundamental and interestingly simple relations between the $\boldsymbol{J}$-, $\boldsymbol{L}$-, and $M$-integrals of straight dislocations are derived revealing at the same time the inherent relation between these integrals. Throughout this paper, interesting similarities between the presenting results for straight dislocations with results for plane cracks are pointed out and discussed.

The paper is organized into nine sections. The basic framework of incompatible linear elasticity of dislocations is briefly presented and reviewed in Section 2. In Section 3, we provide the general expressions of the $\boldsymbol{J}$-, $M$-, and $\boldsymbol{L}$-integrals for dislocations (Lazar and Kirchner, 2007a) and we further discuss their physical meaning and interpretation in the framework of configurational or Eshelbian mechanics. Section 4 is devoted to the derivation of the $\boldsymbol{J}$-, $M$-, and $\boldsymbol{L}$-integrals of straight dislocations valid also for anisotropic materials. In the subsections 4.1–4.3, the explicit formulas of the $\boldsymbol{J}$-, $M$-, and $L_3$-integrals are derived for parallel screw and parallel edge dislocations in $x$-direction as well as in $y$-direction in isotropic materials. Section 5 provides the fundamental relations between the $\boldsymbol{J}$-, $M$-, and $L_3$-integrals of straight dislocations. In Section 6, the examined integrals are directly connected with the interaction energy. In Section 7, the translational as well as the rotational energy-release of straight dislocations are defined and are connected to the $\boldsymbol{J}$-, and $L_3$-integrals, respectively. In Section 8, a stability criterion for straight edge dislocations in terms of the $J_k$-integral is provided and some examples are studied. The main conclusions concerning the $\boldsymbol{J}$-, $M$-, and $L_3$-integrals of straight dislocations are gathered in the last Section.

---

[2]In this work, we use the standard definition of the pre-logarithmic energy factor given in the literature of dislocation theory (see, e.g., Hirth and Lothe (1982); Teodosiu (1982)) and it corresponds to the "dislocation energy factor" used in Rice (1985).

## 2. Preliminaries of dislocation theory

The subject of the present work relies on the framework of incompatible linear elasticity theory. We give below the fundamental equations of three-dimensional incompatible elasticity in the presence of dislocations. When considering a homogeneous linear elastic body containing dislocations, then the elastic energy density reads

$$W = \frac{1}{2} C_{ijkl} \beta_{ij} \beta_{kl} , \tag{1}$$

where $\beta_{ij}$ is the *elastic distortion tensor* and $C_{ijkl}$ is the *tensor of the elastic constants* possessing the usual symmetries $C_{ijkl} = C_{klij} = C_{ijlk} = C_{jikl}$. The stress tensor is defined by

$$\sigma_{ij} = \frac{\partial W}{\partial \beta_{ij}} = C_{ijkl} \beta_{kl} , \tag{2}$$

which is the Hooke law for full anisotropy. Using Eq. (2), the elastic energy density (1) can be written in the form

$$W = \frac{1}{2} \sigma_{ij} \beta_{ij} . \tag{3}$$

For dislocations, which cause self-stresses, the equilibrium condition reads

$$\sigma_{ij,j} = 0 . \tag{4}$$

In the theory of incompatible elasticity, the *total distortion tensor* $\beta_{ij}^{\mathrm{T}}$, which is defined as the gradient of the displacement vector $u_i$, is decomposed into elastic and plastic parts (e.g., Kröner (1958, 1981); Mura (1987))

$$\beta_{ij}^{\mathrm{T}} := u_{i,j} = \beta_{ij} + \beta_{ij}^{\mathrm{P}} , \tag{5}$$

where $\beta_{ij}^{\mathrm{P}}$ is the *plastic distortion tensor* or *eigendistortion*. The incompatibility condition defines the *dislocation density tensor* $\alpha_{ij}$ which is given by (see, e.g., Kröner (1958, 1981))

$$\alpha_{ij} = -\epsilon_{jkl} \beta_{il,k}^{\mathrm{P}} \qquad \text{or} \qquad \alpha_{ij} = \epsilon_{jkl} \beta_{il,k} , \tag{6}$$

where $\epsilon_{jkl}$ is the Levi-Civita tensor. The dislocation density tensor satisfies the Bianchi identity

$$\alpha_{ij,j} = 0 , \tag{7}$$

which means that dislocations cannot end inside the body.

## 3. The general expressions of the *J*-, *M*-, and *L*-integrals for dislocations

In this section, we give the general expressions of the *J*-, *M*-, and *L*-integrals for dislocations as they have been derived by Lazar and Kirchner (2007a) in the framework of incompatible linear elasticity, and we further discuss their physical meaning and interpretation. The *J*-, *M*-, and *L*-integrals correspond to the translational, rotational, and scaling material balance laws, respectively. Let us take an arbitrary infinitesimal functional derivative $\delta U$ of the elastic energy $U = \int_V W \, \mathrm{d}V$, where $V$ is an arbitrary volume of the elastic body containing dislocations. Using Eq. (3), we obtain

$$\delta U = \delta \int_V W \, \mathrm{d}V = \int_V \sigma_{ij} [\delta \beta_{ij}] \, \mathrm{d}V . \tag{8}$$

### 3.1. *J*-integral and Peach-Koehler force

*Material* or *configurational forces* can be obtained by specifying the functional derivative to be translational

$$\delta = (\delta x_k)\partial_k \,, \tag{9}$$

where $(\delta x_k)$ is an infinitesimal shift in the $x_k$-direction. The ***J**-integral of dislocations* is given by (see Kirchner (1999); Lazar and Kirchner (2007a) for technical details)

$$J_k := \int_V \partial_j \big[W\delta_{jk} - \sigma_{ij}\beta_{ik}\big]\mathrm{d}V = \int_V \epsilon_{kjl}\sigma_{ij}\alpha_{il}\,\mathrm{d}V \,. \tag{10}$$

Note that Eq. (10) represents the vectorial ***J***-integral. The quantity in the divergence in the volume integral on the left hand side in Eq. (10) is the *Eshelby stress tensor of dislocations* in incompatible elasticity theory

$$P_{kj} = W\delta_{jk} - \sigma_{ij}\beta_{ik} \,, \tag{11}$$

which has the property that

$$P_{kk} = W \,, \tag{12}$$

since $\delta_{kk} = d = 3$, where $d$ is the space dimension. The integrand of the volume integral on the right hand side in Eq. (10) is nothing but the *Peach-Koehler force density*

$$f_k^{\mathrm{PK}} = \epsilon_{kjl}\sigma_{ij}\alpha_{il} \,, \tag{13}$$

which is the material or configurational force density of a dislocation $\alpha_{il}$ interacting with a stress field $\sigma_{ij}$ in an elastic material. The stress field $\sigma_{ij}$ may be caused by another defect or by an external source (e.g. external forces). Finally, the ***J***-integral has the physical meaning of the *Peach-Koehler force*

$$J_k = \int_V \epsilon_{kjl}\sigma_{ij}\alpha_{il}\,\mathrm{d}V = \mathcal{F}_k^{\mathrm{PK}} \,. \tag{14}$$

Observe that for the calculation of the ***J***-integral using the expression (14) only the dislocation density tensor and the stress tensor of a certain dislocation problem are needed. The physical structure and interpretation of the ***J***-integral as the Peach-Koehler force is visible in the volume integral expression (14). Moreover, the volume integral expression (14) is also used and needed for the simulations in applications like in the so-called dislocation dynamics (see, e.g., Po *et al.* (2014)). Note that in the framework of fracture mechanics for the study of cracks (in the framework of compatible elasticity), the expression of the ***J***-integral as a surface integral of the corresponding Eshelby stress tensor (Maugin, 1993; Kienzler and Herrmann, 2000) is used for the calculation of the ***J***-integral of cracks. One should keep in mind that the ***J***-integral for cracks is path-independent in an otherwise homogeneous material in absence of external forces. There are certain important differences between the treatment of a crack problem and a dislocation problem that are always necessary to be considered.

In the framework of defect-free compatible elasticity, the ***J***-integral is zero, ***J*** = 0, under also the original assumption of absent external forces in a homogeneous material.

### 3.2. *M*-integral and configurational work

In this subsection, we consider *material* or *configurational work* and must, therefore, use the dilatational functional derivative

$$\delta = x_k\partial_k \,. \tag{15}$$

The general expression of the $M$-integral of dislocations valid for a $d$-dimensional problem ($d \geq 2$) is given by (see Lazar and Kirchner (2007a) for technical details)

$$M := \int_V \partial_j \Big[ x_k P_{kj} - \frac{d-2}{2} u_k \sigma_{kj} \Big] \mathrm{d}V = \int_V \Big\{ x_k f_k^{\mathrm{PK}} - \frac{d-2}{2} \beta_{ij}^{\mathrm{P}} \sigma_{ij} \Big\} \mathrm{d}V \,. \tag{16}$$

The quantity in the divergence in the volume integral on the left hand side in Eq. (16) is the *scaling flux vector of dislocations* in incompatible elasticity theory

$$Y_j = x_k P_{kj} - \frac{d-2}{2} u_k \sigma_{kj} \,, \tag{17}$$

where $\delta_{kk} = d$. In field theory, the pre-factor $-(d-2)/2$ is called the *scaling* or *canonical dimension* of the vector field $u_k$ (see also Lazar and Kirchner (2007a)).

In order to find the $M$-integral for dislocations, one has to recall that a dislocation problem is by definition a three-dimensional problem, since a dislocation is a line defect in a three-dimensional crystal. Therefore, the space-dimension $d = 3$ has to be used. Even a problem of straight dislocations (in $z$-direction) is a three-dimensional problem with special symmetries. In this case, the corresponding deformation is a generalized plane strain deformation (plane strain and/or anti-plane strain) where all the three components of the displacement vector $u_i(x, y)$ with $i = x, y, z$ depend only on $x$ and $y$ and the corresponding stresses are calculated by using the three-dimensional Hooke law. Only for two-dimensional materials like graphene or thin films, the space-dimension $d = 2$ has to be used, since for such materials a dislocation "reduces" to a "point defect" (see, e.g., Lazar (2013)). Therefore, for $d = 3$, Eq. (16) gives the $M$-integral of dislocations

$$M = \int_V \partial_j \Big[ x_k P_{kj} - \frac{1}{2} u_k \sigma_{kj} \Big] \mathrm{d}V = \int_V \Big\{ x_k f_k^{\mathrm{PK}} - \frac{1}{2} \beta_{ij}^{\mathrm{P}} \sigma_{ij} \Big\} \mathrm{d}V \,. \tag{18}$$

It can be seen in Eq. (18) that the $M$-integral of dislocations as it is expressed by the volume integral on the right hand side consists of two distinct parts. The first term is the configurational work produced by the Peach-Koehler force density and we can recognize that the second term is actually the dislocation energy caused by a plastic distortion $\beta_{ij}^{\mathrm{P}}$ of a dislocation in the stress field $\sigma_{ij}$ of another dislocation. Hence, the $M$-integral of dislocations can be written as

$$M = \int_V \Big\{ x_k f_k^{\mathrm{PK}} - \frac{1}{2} \beta_{ij}^{\mathrm{P}} \sigma_{ij} \Big\} \mathrm{d}V = \mathcal{W}^{\mathrm{PK}} + U_{\mathrm{d}} \,, \tag{19}$$

where the *configurational work produced by the Peach-Koehler force density* is denoted by

$$\mathcal{W}^{\mathrm{PK}} = \int_V x_k f_k^{\mathrm{PK}} \, \mathrm{d}V \tag{20}$$

and $U_{\mathrm{d}}$ is the *dislocation energy* defined by

$$U_{\mathrm{d}} := -\frac{1}{2} \int_V \beta_{ij}^{\mathrm{P}} \sigma_{ij} \, \mathrm{d}V \,. \tag{21}$$

Note that the dislocation energy is identical to "the elastic strain energy of a dislocation caused by a plastic distortion $\beta_{ij}^{\mathrm{P}}$" as defined by Mura (1987) in the framework of eigenstrain theory. It is derived from the elastic strain energy if the equilibrium condition and free-surface conditions are fulfilled. For reasons of brevity and clarity, the short term dislocation energy is herein used for $U_{\mathrm{d}}$. In addition, the *interaction energy* $U_{\mathrm{int}}$ between a dislocation with plastic distortion $\beta_{ij}^{\mathrm{P}}$ and another dislocation with stress field $\sigma_{ij}$ is defined by (see, e.g., Mura (1969, 1987))

$$U_{\mathrm{int}} = 2 U_{\mathrm{d}} = -\int_V \beta_{ij}^{\mathrm{P}} \sigma_{ij} \, \mathrm{d}V \,. \tag{22}$$

Using the definition (22), the $M$-integral of dislocations can be alternatively written as

$$M = \int_V \left\{ x_k f_k^{\mathrm{PK}} - \frac{1}{2} \beta_{ij}^{\mathrm{P}} \sigma_{ij} \right\} \mathrm{d}V = \mathcal{W}^{\mathrm{PK}} + \frac{1}{2} U_{\mathrm{int}} \,. \tag{23}$$

Analogously to the $\boldsymbol{J}$-integral, it can been seen that Eq. (19), giving the $M$-integral of dislocations as a volume integral where only the plastic distortion tensor, the stress tensor and the Peach-Koehler force density vector are needed for a certain dislocation problem, represents the physical and straightforward way for the calculation of the $M$-integral of dislocations.

In the exceptional case of two-dimensional bodies, that is for $d = 2$, it is interesting to see that the second term of the volume integrals in Eq. (16) vanishes. Then, in the expression of the $M$-integral (Eq. (16) or Eq. (18)), it remains only the contribution of the configurational work $\mathcal{W}^{\mathrm{PK}}$ produced by the Peach-Koehler force density. It should be emphasized that the contribution of the dislocation energy $U_{\mathrm{d}}$ (which is actually the direct contribution of the plastic fields) is rather significant and plays an important role in the derivation and physical interpretation of the $M$-, and $\boldsymbol{L}$-integrals of dislocations in three-dimensional materials as we will see in the next sections.

Note that in the framework of defect-free compatible elasticity, the $M$-integral becomes zero, $M = 0$, under also the original assumption of absent external forces in a homogeneous material.

### 3.3. $\boldsymbol{L}$-integral and configurational vector moments

*Material* or *configurational vector moments* can be derived by taking the rotational functional derivative

$$\delta = (\delta x_k)\epsilon_{kji} x_j \partial_i \,, \tag{24}$$

where $(\delta x_k)$ denotes the $x_k$-direction of the axis of rotation. The $\boldsymbol{L}$-*integral of dislocations* reads (see Lazar and Kirchner (2007a) for technical details)

$$L_k := \int_V \partial_l \big(\epsilon_{kji}[x_j P_{il} + u_j \sigma_{il}]\big) \mathrm{d}V = \int_V \epsilon_{kji} \Big\{ x_j f_i^{\mathrm{PK}} + \beta_{jl}^{\mathrm{P}} \sigma_{il} + [\beta_{jl}\sigma_{il} + \beta_{lj}\sigma_{li}] \Big\} \mathrm{d}V \,. \tag{25}$$

Eq. (25) represents the vectorial $\boldsymbol{L}$-integral. The quantity in the divergence in the volume integral on the left hand side in Eq. (25) is the *total angular momentum tensor of dislocations* in incompatible elasticity theory

$$M_{kl} = \epsilon_{kji}\big[x_j P_{il} + u_j \sigma_{il}\big] \,. \tag{26}$$

The first part in Eq. (26) is the *orbital angular momentum tensor* given in terms of the Eshelby stress tensor (11) and the second part is the *spin angular momentum tensor* for the displacement vector. The volume integral on the right hand side in Eq. (25) consists of three distinct terms. The first term is the material vector moment of the Peach-Koehler force density $f_i^{\mathrm{PK}}$ given in Eq. (13). The second term is the material vector moment caused by the plastic distortion $\beta_{jl}^{\mathrm{P}}$ in presence of the stress field $\sigma_{il}$ and the third term is the material vector moment due to the anisotropy of the material. Similarly to the $\boldsymbol{J}$-, and $M$-integrals, the expression of the $\boldsymbol{L}$-integral as a volume integral on the right hand side in Eq. (25) represents the suitable form of the $\boldsymbol{L}$-integral for dislocations where for example for an isotropic material only the relevant dislocation fields have to be substituted for a certain dislocation problem. Therefore, the $\boldsymbol{L}$-*integral of dislocations for an anisotropic material* is

$$L_k = \int_V \epsilon_{kji} \Big\{ x_j f_i^{\mathrm{PK}} + \beta_{jl}^{\mathrm{P}} \sigma_{il} + [\beta_{jl}\sigma_{il} + \beta_{lj}\sigma_{li}] \Big\} \mathrm{d}V \,, \tag{27}$$

and the $\boldsymbol{L}$-*integral of dislocations for an isotropic material* reduces to

$$L_k = \int_V \epsilon_{kji} \Big\{ x_j f_i^{\mathrm{PK}} + \beta_{jl}^{\mathrm{P}} \sigma_{il} \Big\} \mathrm{d}V \,. \tag{28}$$

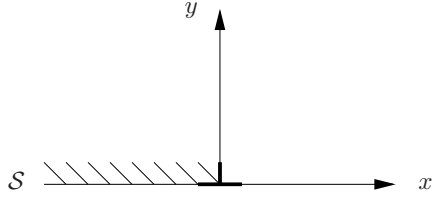

Figure 1: Geometry of a straight dislocation at position $(0,0)$.

Note that in the framework of defect-free compatible elasticity, the **L**-*integral for an anisotropic material* is

$$L_k = \int_V \epsilon_{kji}\Big\{u_{j,l}\sigma_{il} + u_{l,j}\sigma_{li}\Big\} \mathrm{d}V \,, \tag{29}$$

and the **L**-*integral for an isotropic material* is zero, $\boldsymbol{L} = 0$, under also the original assumption of absent external forces in a homogeneous material.

It is obvious that the **J**-, $M$-, and **L**-integrals of dislocations given by Eqs. (14), (19) and (27), respectively, are not conserved integrals since they are, in general, non-zero due to the contributions of the dislocation density and plastic distortion tensors. We conclude that Eqs. (14), (19) and (27) are the appropriate expressions of the **J**-, $M$-, and **L**-integrals of dislocations in anisotropic materials, respectively. The appropriate expressions of the **J**-, $M$-, and **L**-integrals of dislocations in isotropic materials are given by the Eqs. (14), (19) and (28), respectively. These expressions can also be used for solving problems in dislocation-based fracture mechanics (Weertman, 1996).

## 4. $J$-, $M$-, and $L_3$-integrals for straight Volterra dislocations

In this section, we determine the **J**-, $M$-, and **L**-integrals of a straight dislocation at the position $(\bar{x}, \bar{y})$ in the stress field of another straight dislocation at the origin of the coordinate system $(0,0)$. This is a problem of generalized plane strain deformation.

For a straight dislocation, the dislocation line $\mathcal{C}$ is a straight line and the dislocation surface $\mathcal{S}$ is a semi-infinite plane bounded by $\mathcal{C}$. In particular, let $\mathcal{C}$ run along the $z$-axis and $\mathcal{S}$ be the part of the $xz$-plane for negative $x$ (see Fig. 1). On the surface $\mathcal{S}$, the displacement vector $\boldsymbol{u}$ has a jump $\boldsymbol{b}$, which is the Burgers vector. Then the only non-vanishing components of the plastic distortion and dislocation density tensors for a straight Volterra dislocation at position $(\bar{x}, \bar{y})$ are (see also deWit (1973))

$$\beta^{\mathrm{P}}_{iy}(x-\bar{x}, y-\bar{y}) = b_i\, H(-(x-\bar{x}))\, \delta(y-\bar{y})$$
$$= b_i\, \delta(y-\bar{y}) \int_{-\infty}^{0} \delta(x-\bar{x}-x')\, \mathrm{d}x' \,, \tag{30}$$
$$\alpha_{iz}(x-\bar{x}, y-\bar{y}) = b_i\, \delta(x-\bar{x})\, \delta(y-\bar{y}) \,, \tag{31}$$

where $\delta(.)$ and $H(.)$ denote the Dirac delta function and the Heaviside step function, respectively. The plastic distortion $\beta^{\mathrm{P}}_{iy}$ possesses a discontinuity on $\mathcal{S}$.

If we substitute Eqs. (30) and (31) into Eqs. (14), (19), (27) and (28) and perform the (three-dimensional)

volume integration, we get the $\boldsymbol{J}$-, $M$-, and $\boldsymbol{L}$-integrals of straight Volterra dislocations

$$J_k = \mathcal{F}_k^{\text{PK}} = \epsilon_{kjz} b_i \sigma_{ij}(\bar{x}, \bar{y}) \, l_z \,, \tag{32}$$

$$M = \mathcal{W}^{\text{PK}} + U_{\text{d}} = \bar{x}_k J_k - \frac{l_z}{2} \int_{-\infty}^{0} b_i \sigma_{iy}(\bar{x} + x', \bar{y}) \, \mathrm{d}x' \,, \tag{33}$$

$$L_k = \epsilon_{kji} \bar{x}_j J_i + l_z \int_{-\infty}^{0} \epsilon_{kji} b_j \sigma_{iy}(\bar{x} + x', \bar{y}) \, \mathrm{d}x' + \int_V \epsilon_{kji} \big(\beta_{jl} \sigma_{il} + \beta_{lj} \sigma_{li}\big) \, \mathrm{d}V \,, \tag{34}$$

$$L_k = \epsilon_{kji} \bar{x}_j J_i + l_z \int_{-\infty}^{0} \epsilon_{kji} b_j \sigma_{iy}(\bar{x} + x', \bar{y}) \, \mathrm{d}x' \,. \tag{35}$$

Here, $l_z$ denotes the length of the dislocation line and it comes from the performance of the $z$-integration. Eqs. (32)–(34) are valid for straight dislocations in anisotropic materials. Eq. (35) is valid for straight dislocations only in isotropic materials. In Eq. (32), we recover from the general expression of the $\boldsymbol{J}$-integral of dislocations the original Peach-Koehler force formula derived by Peach and Koehler (1950) for straight dislocations. It can be seen in Eqs. (33)–(35) that the first terms appear due to the contribution of the Peach-Koehler force (or $\boldsymbol{J}$-integral) and the second terms appear due to the plastic distortion of a straight dislocation. In Eq. (34), an additional term (volume integral) appears due to the anisotropy of the material. Note that in Eqs. (33)–(35), the position vector $\bar{x}_k = (\bar{x}, \bar{y})$ arises from the position of the dislocation density (31) due to the volume integration. The *dislocation energy for straight dislocations* reads

$$U_{\text{d}} = \frac{1}{2} U_{\text{int}} = -\frac{l_z}{2} \int_{-\infty}^{0} b_i \sigma_{iy}(\bar{x} + x', \bar{y}) \, \mathrm{d}x' \,. \tag{36}$$

Therefore, the *$M$-integral (33) of straight dislocations* can be elegantly written in terms of the $\boldsymbol{J}$-integral and the dislocation energy

$$M = \bar{x}_k J_k + U_{\text{d}} \,, \tag{37}$$

or equivalently in terms of the Peach-Koehler force and the dislocation energy

$$M = \bar{x}_k \mathcal{F}_k^{\text{PK}} + U_{\text{d}} \,. \tag{38}$$

For two-dimensional ($d = 2$) dislocation problems, the $M$-integral (37) reduces to

$$M = \bar{x}_k J_k \,. \tag{39}$$

**Remark 1.** It is remarkable to see that the $M$-integral for cracks in the framework of compatible elasticity in two-dimensions is given by a similar formula to Eq. (39) where $J_k$ is now the corresponding $J_k$-integral for cracks. An analogous relation was found by Freund (1978) for the study of plane cracks in "two-dimensional bodies" where the coordinates are chosen such that the singular point of the stress field is at $x_k^0$. Here, $\bar{x}_k = x_k^0$. Therefore,

$$M = x_1^0 J_1 + x_2^0 J_2 \,. \tag{40}$$

The relation (40) is known and used in the literature of fracture mechanics (see, e.g., Golebiewska-Herrmann and Herrmann (1981b); Kienzler and Herrmann (2000)) in the form

$$M = 2aJ_1 \tag{41}$$

for the study of plane cracks of length $2a$.

**Remark 2.** In the framework of three-dimensional compatible elasticity for isotropic materials, the $\boldsymbol{L}$-integral (35) reduces to

$$L_k = \epsilon_{kji} \bar{x}_j J_i \,, \tag{42}$$

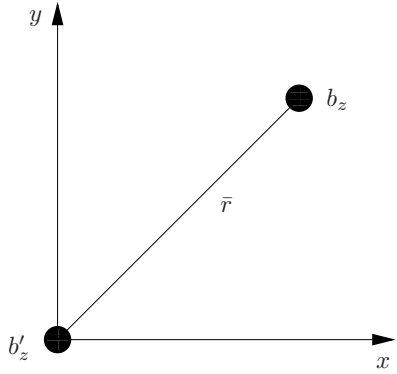

Figure 2: Interaction between two parallel screw dislocations with Burgers vectors in $z$-direction.

where $J_i$ is the corresponding $J_i$-integral of compatible elasticity. It can be seen from Eq. (42) that the $\boldsymbol{L}$-integral is expressed only in terms of the $\boldsymbol{J}$-integral for a problem of a generalized plane strain deformation. In two-dimensional compatible elasticity, using $\bar{x}_1 = 2a$ and $\bar{x}_2 = 0$, Eq. (42) simplifies to

$$L_3 = 2aJ_2 \,. \tag{43}$$

**Remark 3.** In fracture mechanics, a similar relation to Eq. (43) can be found in the literature (see, e.g., Kienzler and Herrmann (2000)) for the study of plane cracks of length $2a$ but derived under special applied stress conditions.

### 4.1. Parallel screw dislocations

Consider two parallel screw dislocations at positions $(\bar{x}, \bar{y})$ and $(0, 0)$ with Burgers vectors $b_z$ and $b'_z$, respectively, in an isotropic material (see Fig. 2). The non-vanishing components of the stress tensor of a screw dislocation with Burgers vector $b_z$ in isotropic elasticity read (deWit, 1973; Landau and Lifschitz, 1986)

$$\sigma_{zx}(x, y) = -\frac{\mu b_z}{2\pi} \frac{y}{x^2 + y^2}, \qquad \sigma_{zy}(x, y) = \frac{\mu b_z}{2\pi} \frac{x}{x^2 + y^2}, \tag{44}$$

where $\mu$ is the shear modulus.

#### 4.1.1. $\boldsymbol{J}$-integral

Using Eq. (44), the $J_1$- and $J_2$-integrals of the two parallel screw dislocations per unit dislocation length $l_z$ are obtained from Eq. (32)

$$\frac{J_1}{l_z} = \frac{J_x}{l_z} = \frac{\mathcal{F}_x^{\text{PK}}}{l_z} = b_z \sigma_{zy}(\bar{x}, \bar{y}) = \frac{\mu b_z b'_z}{2\pi} \frac{\bar{x}}{\bar{x}^2 + \bar{y}^2}, \tag{45}$$

$$\frac{J_2}{l_z} = \frac{J_y}{l_z} = \frac{\mathcal{F}_y^{\text{PK}}}{l_z} = -b_z \sigma_{zx}(\bar{x}, \bar{y}) = \frac{\mu b_z b'_z}{2\pi} \frac{\bar{y}}{\bar{x}^2 + \bar{y}^2}. \tag{46}$$

As it can be seen in Eqs. (45) and (46), the $J_1$- and $J_2$-integrals give the two non-vanishing components of the Peach-Koehler force between the two screw dislocations and depend only on the position $(\bar{x}, \bar{y})$ of the dislocation with Burgers vector $b_z$ which is in the stress field of the screw dislocation with Burgers vector $b'_z$ at the position $(0, 0)$ in an elastic material with the material constant $\mu$.

In cylindrical coordinates, the two components $J_r$ and $J_\varphi$ of the $\boldsymbol{J}$-integral read

$$\frac{J_r}{l_z} = \frac{\mathcal{F}_r^{\text{PK}}}{l_z} = \frac{1}{l_z}(J_x \cos\varphi + J_y \sin\varphi) = \frac{\mu b_z b'_z}{2\pi \bar{r}}, \tag{47}$$

$$\frac{J_\varphi}{l_z} = \frac{\mathcal{F}_\varphi^{\text{PK}}}{l_z} = \frac{1}{l_z}(J_y \cos\varphi - J_x \sin\varphi) = 0, \tag{48}$$

where $\bar{r} = \sqrt{\bar{x}^2 + \bar{y}^2}$ is the distance between the two dislocations (see Fig. 2). Therefore, the Peach-Koehler force between two parallel screw dislocations is a purely radial force. It is repulsive if $b_z b'_z > 0$ and it is attractive if $b_z b'_z < 0$. The radial force cannot be zero and hence no stable configurations are formed by parallel screw dislocations.

*4.1.2. M-integral*

For the problem under consideration, the $M$-integral (33) reads

$$M = \mathcal{W}^{\text{PK}} + U_{\text{d}} = \bar{x} J_x + \bar{y} J_y - \frac{l_z}{2} \int_{-\infty}^{0} b_z \, \sigma_{zy}(\bar{x} + x', \bar{y}) \, \mathrm{d}x' \,. \tag{49}$$

Substituting the $J_1$-integral (45), the $J_2$-integral (46) and the stress field $\sigma_{zy}$ from Eq. (44) into Eq. (49), the $M$-integral per unit dislocation length is given by

$$\frac{M}{l_z} = \frac{\mathcal{W}^{\text{PK}}}{l_z} + \frac{U_{\text{d}}}{l_z} = \frac{\mu b_z b'_z}{2\pi} - \frac{\mu b_z b'_z}{4\pi} \int_{-\infty}^{0} \frac{\bar{x} + x'}{(\bar{x} + x')^2 + \bar{y}^2} \, \mathrm{d}x' \,. \tag{50}$$

We carry out the integral part in Eq. (50)

$$\int_{-\infty}^{0} \frac{\bar{x} + x'}{(\bar{x} + x')^2 + \bar{y}^2} \, \mathrm{d}x' = \frac{1}{2} \ln \left( (\bar{x} + x')^2 + \bar{y}^2 \right) \Big|_{-\infty}^{0} = \ln \frac{\bar{r}}{L} \,, \tag{51}$$

where the limit to infinity is replaced by a finite number $L$, which corresponds to the size of the dislocated body. Using Eq. (51), Eq. (50) reads

$$\frac{M}{l_z} = \frac{\mathcal{W}^{\text{PK}}}{l_z} + \frac{U_{\text{d}}}{l_z} = \frac{\mu b_z b'_z}{2\pi} - \frac{\mu b_z b'_z}{4\pi} \ln \frac{\bar{r}}{L} \,. \tag{52}$$

In Eq. (52), we can recognize the so-called *pre-logarithmic energy factor* (e.g., Hirth and Lothe (1982); Teodosiu (1982)) for screw dislocations

$$K^s_{zz} = \frac{\mu b_z b'_z}{4\pi} \,. \tag{53}$$

Thus, the *configurational work* $\mathcal{W}^{\text{PK}}$ done by the Peach-Koehler force between two parallel screw dislocations per unit dislocation length equals twice the pre-logarithmic energy factor

$$\frac{\mathcal{W}^{\text{PK}}}{l_z} = \frac{\bar{x}_k J_k}{l_z} = \frac{\mu b_z b'_z}{2\pi} = 2 K^s_{zz} \,, \tag{54}$$

which is constant. The *dislocation energy of two parallel screw dislocations* per unit dislocation length reads

$$\frac{U_{\text{d}}}{l_z} = -\frac{\mu b_z b'_z}{4\pi} \ln \frac{\bar{r}}{L} = -K^s_{zz} \ln \frac{\bar{r}}{L} \,, \tag{55}$$

which is a logarithmic term depending only on the distance $\bar{r}$ between the two screw dislocations for a given extension $L$ of the crystal. Note that the dislocation energy $U_{\text{d}}$ is produced by the plastic distortion of the screw dislocation at the position $(\bar{x}, \bar{y})$ with Burgers vector $b_z$ in the stress field of the screw dislocation at the position $(0, 0)$ with Burgers vector $b'_z$ (see Eq. (21)). The final result (Eq. (52)) for the $M$-integral of *two parallel screw dislocations* per unit dislocation length is

$$\frac{M}{l_z} = \frac{\mu b_z b'_z}{4\pi} \left[ 2 - \ln \frac{\bar{r}}{L} \right] \,, \tag{56}$$

or in terms of the pre-logarithmic energy factor

$$\frac{M}{l_z} = K^s_{zz} \left[ 2 - \ln \frac{\bar{r}}{L} \right] \,. \tag{57}$$

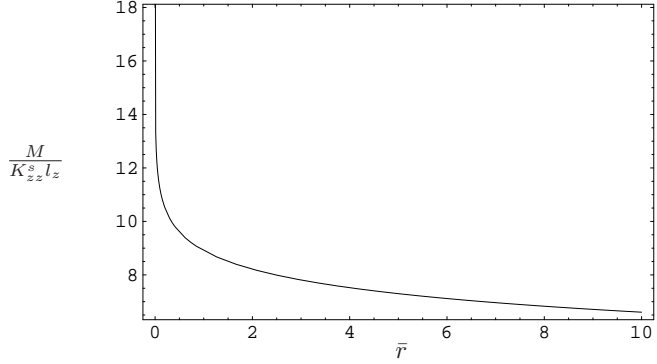

Figure 3: Normalized $M$-integral of two parallel screw dislocations for $b_z b'_z > 0$ against the distance $\bar{r}$ with $L = 1000$.

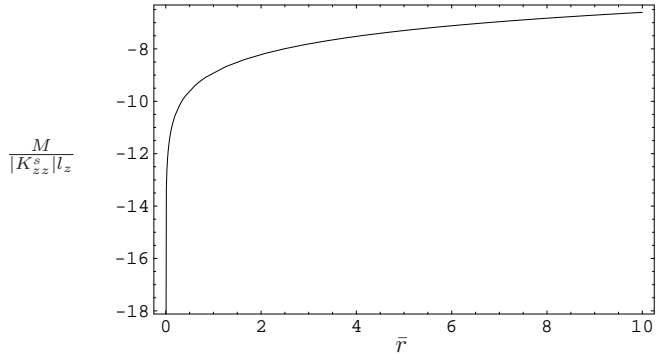

Figure 4: Normalized $M$-integral of two parallel screw dislocations for $b_z b'_z < 0$ against the distance $\bar{r}$ with $L = 1000$.

We need to emphasize here that the logarithmic term depending on the distance $\bar{r}$ between the two screw dislocations is of major physical importance, otherwise the $M$-integral would be constant.

Recalling Eq. (23), the *$M$-integral (56) of two parallel screw dislocations in terms of the corresponding interaction energy* can be written as follows

$$\frac{M}{l_z} = 2K^s_{zz} + \frac{1}{2}\frac{U_{\text{int}}}{l_z}, \tag{58}$$

where

$$\frac{U_{\text{int}}}{l_z} = -2K^s_{zz} \ln \frac{\bar{r}}{L}. \tag{59}$$

Eq. (58) states that the $M$-integral of two parallel screw dislocations in statics (per unit dislocation length) can be interpreted as interaction energy, since it is the sum of the half of the interaction energy $U_{\text{int}}$ between the two dislocations (per unit dislocation length) plus a constant term (namely twice the pre-logarithmic energy factor). To sum up, the constant term in Eq. (58) emanates from the Peach-Koehler force or $\boldsymbol{J}$-integral and the most important contribution, the logarithmic term emanates from the dislocation energy (direct contribution of the plastic fields) and gives the physical meaning to the $M$-integral.

The normalized $M$-integral of two parallel screw dislocations is plotted in Figs. 3 and 4. In Fig. 3, it can be seen that for parallel screw dislocations with Burgers vectors $b_z$ and $b'_z$ of the same sign ($b_z b'_z > 0$) when $\bar{r}$ increases the $M$-integral decreases. If the Burgers vectors are of opposite sign ($b_z b'_z < 0$), then the $M$-integral is increasing when $\bar{r}$ increases (see Fig. 4).

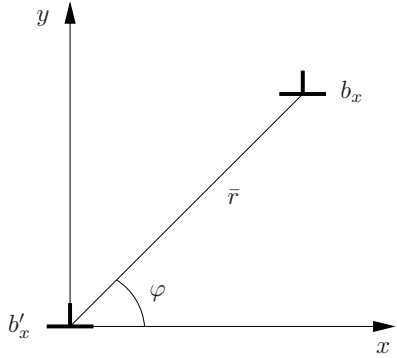

Figure 5: Interaction between two parallel edge dislocations with Burgers vectors in $x$-direction.

Observe that the non-vanishing component $J_r$ of the $\boldsymbol{J}$-integral (see Eq. (47)) can also be expressed in terms of the pre-logarithmic energy factor $K^s_{zz}$ as follows

$$\frac{J_r}{l_z} = \frac{\mathcal{F}^{\text{PK}}_r}{l_z} = 2K^s_{zz}\frac{1}{\bar{r}}\,. \tag{60}$$

#### 4.1.3. $L_3$-integral

The $L_3$-integral for an isotropic material, Eq. (35), reads for screw dislocations

$$L_3 = L_z = \epsilon_{zji}\bar{x}_j J_i = \bar{x}J_y - \bar{y}J_x\,, \tag{61}$$

since the second term in Eq. (35) is zero for screw dislocations. Substituting Eqs. (45) and (46) into Eq. (61), we find

$$L_3 = 0\,, \tag{62}$$

which shows that the $L_3$-integral for screw dislocations in an isotropic material is zero, and therefore, it is a conserved integral. Thus, there is no rotational moment about the $z$-axis (see also subsection 4.2.3) produced by screw dislocations in isotropic materials. The reason is that a screw dislocation possesses a $SO(2)$ (cylindrical) symmetry around the dislocation line, which is in the $z$-direction.

### 4.2. Parallel edge dislocations with Burgers vectors in $x$-direction

Consider two parallel edge dislocations at the positions $(\bar{x}, \bar{y})$ and $(0, 0)$ with parallel Burgers vectors $b_x$ and $b'_x$, respectively, in an isotropic material (see Fig. 5). The non-vanishing components of the stress tensor of an edge dislocation with Burgers vector $b_x$ in isotropic elasticity (deWit, 1973; Landau and Lifschitz, 1986; Hirth and Lothe, 1982) are

$$\sigma_{xx}(x,y) = -b_x D\,\frac{y\,(3x^2+y^2)}{(x^2+y^2)^2}\,, \qquad \sigma_{xy}(x,y) = b_x D\,\frac{x\,(x^2-y^2)}{(x^2+y^2)^2}\,, \tag{63}$$
$$\sigma_{yy}(x,y) = b_x D\,\frac{y\,(x^2-y^2)}{(x^2+y^2)^2}\,, \qquad \sigma_{zz}(x,y) = -2\nu b_x D\,\frac{y}{x^2+y^2}\,,$$

where

$$D = \frac{\mu}{2\pi(1-\nu)} \tag{64}$$

and $\nu$ is the Poisson ratio.

### 4.2.1. *J*-integral

Using Eq. (63), the $J_1$- and $J_2$-integrals of two parallel edge dislocations with Burgers vectors in *x*-direction per unit dislocation length are obtained from Eq. (32)

$$\frac{J_1}{l_z} = \frac{J_x}{l_z} = \frac{\mathcal{F}_x^{\text{PK}}}{l_z} = b_x \sigma_{xy}(\bar{x}, \bar{y}) = b_x b_x' D \frac{\bar{x}(\bar{x}^2 - \bar{y}^2)}{(\bar{x}^2 + \bar{y}^2)^2}, \tag{65}$$

$$\frac{J_2}{l_z} = \frac{J_y}{l_z} = \frac{\mathcal{F}_y^{\text{PK}}}{l_z} = -b_x \sigma_{xx}(\bar{x}, \bar{y}) = b_x b_x' D \frac{\bar{y}(3\bar{x}^2 + \bar{y}^2)}{(\bar{x}^2 + \bar{y}^2)^2}, \tag{66}$$

which give the non-vanishing components of the Peach-Koehler force between the two edge dislocations and they depend only on the position $(\bar{x}, \bar{y})$ of the dislocation with Burgers vector $b_x$ which is in the stress field of the edge dislocation with Burgers vector $b_x'$ at the position $(0,0)$. Physically, $J_1$ is the glide force (in *x*-direction) and $J_2$ is the climb force (in *y*-direction).

In cylindrical coordinates, the two non-vanishing components $J_r$ and $J_\varphi$ of the ***J***-integral are given by

$$\frac{J_r}{l_z} = \frac{\mathcal{F}_r^{\text{PK}}}{l_z} = \frac{1}{l_z}(J_x \cos\varphi + J_y \sin\varphi) = \frac{\mu b_x b_x'}{2\pi(1-\nu)\bar{r}}, \tag{67}$$

$$\frac{J_\varphi}{l_z} = \frac{\mathcal{F}_\varphi^{\text{PK}}}{l_z} = \frac{1}{l_z}(J_y \cos\varphi - J_x \sin\varphi) = \frac{\mu b_x b_x'}{2\pi(1-\nu)\bar{r}} \sin 2\varphi, \tag{68}$$

where $\varphi$ is the location angle between the two edge dislocations (see Fig. 5). Due to the component (68), the Peach-Koehler force between two parallel edge dislocations is not a purely radial force.

### 4.2.2. *M*-integral

For the case of two parallel edge dislocations as it has been described above, the *M*-integral (33) reads

$$M = \mathcal{W}^{\text{PK}} + U_{\text{d}} = \bar{x} J_x + \bar{y} J_y - \frac{l_z}{2} \int_{-\infty}^{0} b_x \sigma_{xy}(\bar{x} + x', \bar{y}) \, dx'. \tag{69}$$

If we substitute the $J_1$-integral (65), the $J_2$-integral (66) and the stress field $\sigma_{xy}$ from Eq. (63) into Eq. (69), then the *M*-integral per unit dislocation length becomes

$$\frac{M}{l_z} = \frac{\mathcal{W}^{\text{PK}}}{l_z} + \frac{U_{\text{d}}}{l_z} = b_x b_x' D - \frac{1}{2} b_x b_x' D \int_{-\infty}^{0} \frac{(\bar{x} + x')((\bar{x} + x')^2 - \bar{y}^2)}{((\bar{x} + x')^2 + \bar{y}^2)^2} \, dx'. \tag{70}$$

We perform the integral in Eq. (70)

$$\int_{-\infty}^{0} \frac{(\bar{x} + x')((\bar{x} + x')^2 - \bar{y}^2)}{((\bar{x} + x')^2 + \bar{y}^2)^2} \, dx' = \left[\frac{1}{2}\ln((\bar{x} + x')^2 + \bar{y}^2) + \frac{\bar{y}^2}{(\bar{x} + x')^2 + \bar{y}^2}\right]\bigg|_{-\infty}^{0}$$

$$= \ln\frac{\bar{r}}{L} + \sin^2\varphi, \tag{71}$$

where in the logarithmic term the limit to infinity is replaced by $L$. Substituting Eq. (71) into Eq. (70), we get

$$\frac{M}{l_z} = \frac{\mathcal{W}^{\text{PK}}}{l_z} + \frac{U_{\text{d}}}{l_z} = \frac{\mu b_x b_x'}{2\pi(1-\nu)} - \frac{\mu b_x b_x'}{4\pi(1-\nu)}\left[\ln\frac{\bar{r}}{L} + \sin^2\varphi\right]. \tag{72}$$

The pre-factor in Eq. (72) is the *pre-logarithmic energy factor* (see, e.g., Hirth and Lothe (1982); Teodosiu (1982)) for edge dislocations with Burgers vectors in *x*-direction

$$K_{xx}^e = \frac{\mu b_x b_x'}{4\pi(1-\nu)}. \tag{73}$$

Thus, the *configurational work* $\mathcal{W}^{\mathrm{PK}}$ *done by the Peach-Koehler force between two parallel edge dislocations with Burgers vectors in x-direction* per unit dislocation length equals twice the pre-logarithmic energy factor

$$\frac{\mathcal{W}^{\mathrm{PK}}}{l_z} = \frac{\bar{x}_k J_k}{l_z} = \frac{\mu b_x b'_x}{2\pi(1-\nu)} = 2K^e_{xx}, \tag{74}$$

which is constant. Eq. (74) is in agreement with the expression given by Lothe (1992) for the work done against the interaction force between two edge dislocations.

The *dislocation energy of two parallel edge dislocations with Burgers vectors in x-direction* per unit dislocation length reads

$$\frac{U_{\mathrm{d}}}{l_z} = -\frac{\mu b_x b'_x}{4\pi(1-\nu)}\left[\ln\frac{\bar{r}}{L} + \sin^2\varphi\right] = -K^e_{xx}\left[\ln\frac{\bar{r}}{L} + \sin^2\varphi\right]. \tag{75}$$

The dislocation energy (75), which is done by the plastic distortion of the first edge dislocation in the stress field of the second edge dislocation, constitutes of two terms; a logarithmic term depending on the distance $\bar{r}$ between the two edge dislocations for a given extension $L$ of the body and a second term depending on the location angle $\varphi$.

The final result (Eq. (72)) for the *M-integral of two parallel edge dislocations with Burgers vectors in x-direction* per unit dislocation length is

$$\frac{M}{l_z} = \frac{\mu b_x b'_x}{4\pi(1-\nu)}\left[2 - \ln\frac{\bar{r}}{L} - \sin^2\varphi\right], \tag{76}$$

or in terms of the pre-logarithmic energy factor $K^e_{xx}$

$$\frac{M}{l_z} = K^e_{xx}\left[2 - \ln\frac{\bar{r}}{L} - \sin^2\varphi\right]. \tag{77}$$

The *M-integral between two edge dislocations with Burgers vectors in x-direction in terms of the corresponding interaction energy* reads

$$\frac{M}{l_z} = 2K^e_{xx} + \frac{1}{2}\frac{U_{\mathrm{int}}}{l_z}, \tag{78}$$

where

$$\frac{U_{\mathrm{int}}}{l_z} = -2K^e_{xx}\left[\ln\frac{\bar{r}}{L} + \sin^2\varphi\right]. \tag{79}$$

Eq. (78) states that the $M$-integral of two parallel edge dislocations with Burgers vectors in $x$-direction per unit dislocation length is the half of the interaction energy between the two dislocations per unit dislocation length, namely between the edge dislocation at the position $(0,0)$ with Burgers vector $b'_x$ acting on the edge dislocation at the position $(\bar{x}, \bar{y})$ with Burgers vector $b_x$, plus twice the pre-logarithmic energy factor $K^e_{xx}$.

The normalized $M$-integral of the considered two edge dislocations against the distance $\bar{r}$ is plotted in Figs. 6 and 7. It can be seen in Fig. 6 that for parallel edge dislocations with Burgers vectors $b_x$ and $b'_x$ of the same sign when $\bar{r}$ increases, then the $M$-integral decreases. If the Burgers vectors are of opposite sign, then the $M$-integral increases when $\bar{r}$ increases (see Fig. 7). The normalized $M$-integral of two parallel edge dislocations with Burgers vectors in $x$-direction against the location angle $\varphi$ is plotted in Fig. 8. At $\varphi = \pi/2$, the $M$-integral (interaction energy) possesses a minimum which corresponds to the stable equilibrium position of these edge dislocations against glide as we will show in the subsection 8.1.

**Remark 4.** It is worth noticing that the graph of the $M$-integral between two edge dislocations plotted in Figs. 6 and 8 is similar to the graph of the $M$-integral between two microcracks in homogeneous brittle solids given by Chen (2001b, 2002).

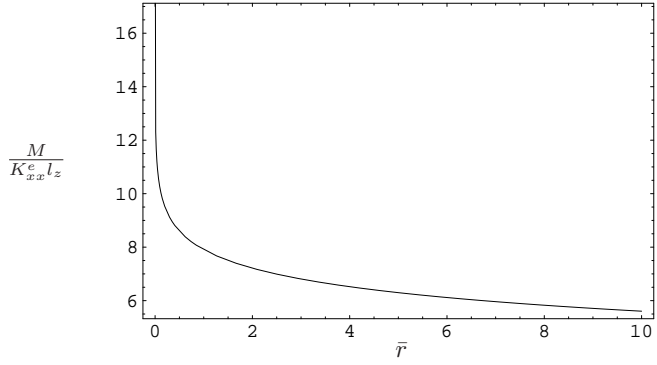

Figure 6: Normalized $M$-integral of two parallel edge dislocations with $b_x b'_x > 0$ against the distance $\bar{r}$ with $\varphi = \pi/2$ and $L = 1000$.

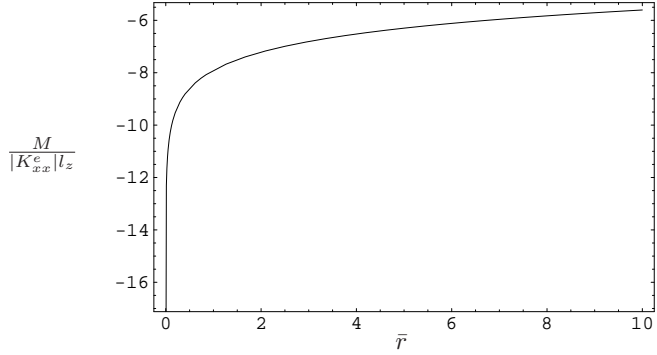

Figure 7: Normalized $M$-integral of two parallel edge dislocations with $b_x b'_x < 0$ against the distance $\bar{r}$ with $\varphi = \pi/2$ and $L = 1000$.

Moreover, the non-vanishing components $J_r$ and $J_\varphi$ of the $\boldsymbol{J}$-integral, Eqs. (67) and (68), respectively, can be written in terms of the pre-logarithmic energy factor (73) as follows

$$\frac{J_r}{l_z} = \frac{\mathcal{F}_r^{\mathrm{PK}}}{l_z} = 2K_{xx}^e \frac{1}{\bar{r}}, \tag{80}$$

$$\frac{J_\varphi}{l_z} = \frac{\mathcal{F}_\varphi^{\mathrm{PK}}}{l_z} = 2K_{xx}^e \frac{\sin 2\varphi}{\bar{r}}. \tag{81}$$

4.2.3. $L_3$-integral

The only non-vanishing component of the $\boldsymbol{L}$-integral (35) reads for parallel edge dislocations with parallel Burgers vectors in $x$-direction

$$L_3 = L_z = \bar{x} J_y - \bar{y} J_x + l_z \int_{-\infty}^{0} b_x \sigma_{yy}(\bar{x} + x', \bar{y}) \, \mathrm{d}x'. \tag{82}$$

Substituting the $J_1$-integral (65), the $J_2$-integral (66) and the stress field $\sigma_{yy}$ from Eq. (63) into Eq. (82), the $L_3$-integral per unit dislocation length becomes

$$\frac{L_3}{l_z} = b_x b'_x D \sin 2\varphi + b_x b'_x D \int_{-\infty}^{0} \frac{\bar{y}((\bar{x} + x')^2 - \bar{y}^2)}{((\bar{x} + x')^2 + \bar{y}^2)^2} \, \mathrm{d}x'. \tag{83}$$

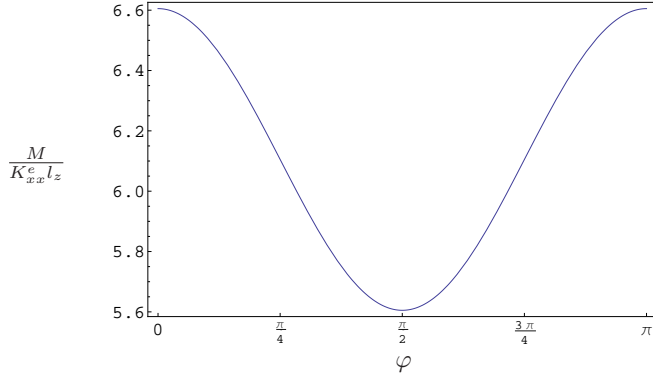

Figure 8: Normalized $M$-integral of two parallel edge dislocations with $b_x b'_x > 0$ against $\varphi$ with $\bar{r}/L = 10/1000$.

We carry out the integral part in Eq. (83)

$$\int_{-\infty}^{0} \frac{\bar{y}((\bar{x}+x')^2 - \bar{y}^2)}{((\bar{x}+x')^2 + \bar{y}^2)^2} \, dx' = -\frac{(\bar{x}+x')\bar{y}}{(\bar{x}+x')^2 + \bar{y}^2}\bigg|_{-\infty}^{0} = -\frac{1}{2}\sin 2\varphi. \tag{84}$$

Substituting Eq. (84) into Eq. (83), we get the $L_3$-*integral of two parallel edge dislocations with Burgers vectors in x-direction* per unit dislocation length

$$\frac{L_3}{l_z} = \frac{\mu b_x b'_x}{4\pi(1-\nu)} \sin 2\varphi, \tag{85}$$

or in terms of the pre-logarithmic energy factor $K_{xx}^e$ (Eq. (73))

$$\frac{L_3}{l_z} = K_{xx}^e \sin 2\varphi. \tag{86}$$

It is worth noticing that both the Peach-Koehler force ($\boldsymbol{J}$-integral) and the plastic distortion give a contribution to the $L_3$-integral, which is of the same character $\sin 2\varphi$. The $L_3$-integral has the physical interpretation of a torque or rotational moment (about the $z$-axis) between the two dislocations. Note that Eq. (85) is in exact agreement with the expression of the "rotating moment due to the dislocation-dislocation interaction" as it is called and given by Lothe (1992).

**Remark 5.** It is interesting to observe that the $L$-integral for cracks and cavities is also a function of $\sin 2\varphi$, as it has been derived by Pak et al. (2012).

It can be seen that the $L_3$-integral is, in general, non-zero and depends only on the angle $\varphi$. $L_3 = 0$ if $\varphi = \{0, \pi/2, \pi\}$.
If $b_x b'_x > 0$, then $L_3$ possesses maximum value at $\varphi = \pi/4$ and minimum value at $\varphi = 3\pi/4$, that is,

$$\max \frac{L_3}{l_z}\left(\varphi = \frac{\pi}{4}\right) = K_{xx}^e, \quad \min \frac{L_3}{l_z}\left(\varphi = \frac{3\pi}{4}\right) = -K_{xx}^e. \tag{87}$$

If $b_x b'_x < 0$, then

$$\max \frac{L_3}{l_z}\left(\varphi = \frac{3\pi}{4}\right) = |K_{xx}^e|, \quad \min \frac{L_3}{l_z}\left(\varphi = \frac{\pi}{4}\right) = -|K_{xx}^e|. \tag{88}$$

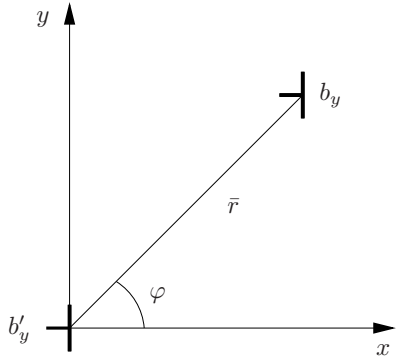

Figure 9: Interaction between two parallel edge dislocations with Burgers vectors in $y$-direction.

*4.3. Parallel edge dislocations with Burgers vectors in y-direction*

Consider two parallel edge dislocations at the positions $(\bar{x}, \bar{y})$ and $(0,0)$ with parallel Burgers vectors $b_y$ and $b'_y$, respectively, in an isotropic material (see Fig. 9). The non-vanishing components of the stress tensor of an edge dislocation with Burgers vector $b_y$ in isotropic elasticity (deWit, 1973; Weertman, 1996) are

$$\sigma_{xx}(x,y) = b_y D \frac{x(x^2 - y^2)}{(x^2+y^2)^2}, \qquad \sigma_{xy}(x,y) = b_y D \frac{y(x^2 - y^2)}{(x^2+y^2)^2}, \qquad (89)$$
$$\sigma_{yy}(x,y) = b_y D \frac{x(x^2 + 3y^2)}{(x^2+y^2)^2}, \qquad \sigma_{zz}(x,y) = 2\nu b_y D \frac{x}{x^2+y^2}.$$

*4.3.1. $\boldsymbol{J}$-integral*

Using Eq. (89), the $J_1$- and $J_2$-integrals of two parallel edge dislocations with Burgers vectors in $y$-direction per unit dislocation length are obtained from Eq. (32)

$$\frac{J_1}{l_z} = \frac{J_x}{l_z} = \frac{\mathcal{F}_x^{\text{PK}}}{l_z} = b_y \sigma_{yy}(\bar{x}, \bar{y}) = b_y b'_y D \frac{\bar{x}(\bar{x}^2 + 3\bar{y}^2)}{(\bar{x}^2+\bar{y}^2)^2}, \qquad (90)$$

$$\frac{J_2}{l_z} = \frac{J_y}{l_z} = \frac{\mathcal{F}_y^{\text{PK}}}{l_z} = -b_y \sigma_{xy}(\bar{x}, \bar{y}) = -b_y b'_y D \frac{\bar{y}(\bar{x}^2 - \bar{y}^2)}{(\bar{x}^2+\bar{y}^2)^2}, \qquad (91)$$

which give the non-vanishing components of the Peach-Koehler force between the two edge dislocations and they depend only on the position $(\bar{x}, \bar{y})$ of the dislocation with Burgers vector $b_y$ which is in the stress field of the edge dislocation with Burgers vector $b'_y$ at the position $(0, 0)$. Notice that now $J_1$ is the climb force (in $x$-direction) and $J_2$ is the glide force (in $y$-direction).

In cylindrical coordinates, the two non-vanishing components $J_r$ and $J_\varphi$ of the $\boldsymbol{J}$-integral are

$$\frac{J_r}{l_z} = \frac{\mathcal{F}_r^{\text{PK}}}{l_z} = \frac{1}{l_z}(J_x \cos\varphi + J_y \sin\varphi) = \frac{\mu b_y b'_y}{2\pi(1-\nu)\bar{r}}, \qquad (92)$$

$$\frac{J_\varphi}{l_z} = \frac{\mathcal{F}_\varphi^{\text{PK}}}{l_z} = \frac{1}{l_z}(J_y \cos\varphi - J_x \sin\varphi) = -\frac{\mu b_y b'_y}{2\pi(1-\nu)\bar{r}} \sin 2\varphi. \qquad (93)$$

*4.3.2. M-integral*

The $M$-integral (33) for the case under study reads

$$M = \mathcal{W}^{\text{PK}} + U_{\text{d}} = \bar{x} J_x + \bar{y} J_y - \frac{l_z}{2} \int_{-\infty}^{0} b_y \sigma_{yy}(\bar{x} + x', \bar{y}) \, \mathrm{d}x'. \qquad (94)$$

If we substitute the $J_1$-integral (90), the $J_2$-integral (91) and the stress field $\sigma_{yy}$ from Eq. (89) into Eq. (94), then the $M$-integral per unit dislocation length reads

$$\frac{M}{l_z} = \frac{\mathcal{W}^{\text{PK}}}{l_z} + \frac{U_{\text{d}}}{l_z} = b_y b'_y D - \frac{1}{2} b_y b'_y D \int_{-\infty}^0 \frac{(\bar{x}+x')((\bar{x}+x')^2 + 3\bar{y}^2)}{((\bar{x}+x')^2 + \bar{y}^2)^2}\,\mathrm{d}x'. \tag{95}$$

We calculate the integral in Eq. (95)

$$\int_{-\infty}^0 \frac{(\bar{x}+x')((\bar{x}+x')^2 + 3\bar{y}^2)}{((\bar{x}+x')^2 + \bar{y}^2)^2}\,\mathrm{d}x' = \left[\frac{1}{2}\ln((\bar{x}+x')^2 + \bar{y}^2) - \frac{\bar{y}^2}{(\bar{x}+x')^2 + \bar{y}^2}\right]\bigg|_{-\infty}^0$$
$$= \ln\frac{\bar{r}}{L} - \sin^2\varphi. \tag{96}$$

Substituting Eq. (96) into Eq. (95), we obtain

$$\frac{M}{l_z} = \frac{\mathcal{W}^{\text{PK}}}{l_z} + \frac{U_{\text{d}}}{l_z} = \frac{\mu b_y b'_y}{2\pi(1-\nu)} - \frac{\mu b_y b'_y}{4\pi(1-\nu)}\left[\ln\frac{\bar{r}}{L} - \sin^2\varphi\right]. \tag{97}$$

The pre-factor in Eq. (97) is the *pre-logarithmic energy factor* for edge dislocations with Burgers vectors in $y$-direction

$$K^e_{yy} = \frac{\mu b_y b'_y}{4\pi(1-\nu)}. \tag{98}$$

Thus, the *configurational work* $\mathcal{W}^{\text{PK}}$ *done by the Peach-Koehler force between two edge dislocations with Burgers vectors in $y$-direction* per unit dislocation length equals twice the pre-logarithmic energy factor

$$\frac{\mathcal{W}^{\text{PK}}}{l_z} = \frac{\bar{x}_k J_k}{l_z} = \frac{\mu b_y b'_y}{2\pi(1-\nu)} = 2K^e_{yy}. \tag{99}$$

The *dislocation energy of two parallel edge dislocations with Burgers vectors in $y$-direction* per unit dislocation length reads

$$\frac{U_{\text{d}}}{l_z} = -\frac{\mu b_y b'_y}{4\pi(1-\nu)}\left[\ln\frac{\bar{r}}{L} - \sin^2\varphi\right] = -K^e_{yy}\left[\ln\frac{\bar{r}}{L} - \sin^2\varphi\right]. \tag{100}$$

The final result (Eq. (97)) for the *$M$-integral of two parallel edge dislocations with Burgers vectors in $y$-direction* per unit dislocation length is

$$\frac{M}{l_z} = \frac{\mu b_y b'_y}{4\pi(1-\nu)}\left[2 - \ln\frac{\bar{r}}{L} + \sin^2\varphi\right], \tag{101}$$

or in terms of the pre-logarithmic energy factor $K^e_{yy}$

$$\frac{M}{l_z} = K^e_{yy}\left[2 - \ln\frac{\bar{r}}{L} + \sin^2\varphi\right]. \tag{102}$$

The mathematical structure and the physical meaning of Eqs. (101) and (102) is similar and analogous to Eqs. (76) and (77), respectively. The only difference is the sign in front of the sin-function.

The *$M$-integral between two edge dislocations with Burgers vectors in $y$-direction in terms of the corresponding interaction energy* reads

$$\frac{M}{l_z} = 2K^e_{yy} + \frac{1}{2}\frac{U_{\text{int}}}{l_z}, \tag{103}$$

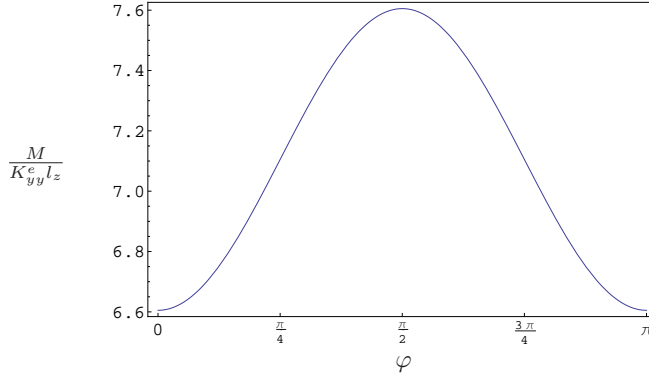

Figure 10: Normalized $M$-integral of two parallel edge dislocations for $b_y b'_y > 0$ against $\varphi$ with $\bar{r}/L = 10/1000$.

where
$$\frac{U_{\text{int}}}{l_z} = -2K^e_{yy}\left[\ln\frac{\bar{r}}{L} - \sin^2\varphi\right]. \tag{104}$$

Eq. (103) states that the $M$-integral of two parallel edge dislocations with Burgers vectors in $y$-direction per unit dislocation length is the half of the interaction energy between the two dislocations per unit dislocation length, namely between the edge dislocation at the position $(0,0)$ with Burgers vector $b'_y$ acting on the edge dislocation at the position $(\bar{x}, \bar{y})$ with Burgers vector $b_y$, plus twice the pre-logarithmic energy factor $K^e_{yy}$.

The normalized $M$-integral of two parallel edge dislocations with Burgers vectors in $y$-direction against the distance $\bar{r}$ is similar to that one of two parallel edge dislocations with Burgers vectors in $x$-direction (see Figs. 6 and 7). The normalized $M$-integral of two edge dislocations with Burgers vectors in $y$-direction against the location angle $\varphi$ is plotted in Fig. 10. The $M$-integral (interaction energy) possesses a minimum at $\varphi = 0$ and $\varphi = \pi$, which correspond to stable equilibrium positions of these edge dislocations as we will show in the subsection 8.2.

In addition, the non-vanishing components $J_r$ and $J_\varphi$ of the $\boldsymbol{J}$-integral, Eqs. (92) and (93), respectively, are written in terms of the pre-logarithmic energy factor (98) as follows
$$\frac{J_r}{l_z} = \frac{\mathcal{F}^{\text{PK}}_r}{l_z} = 2K^e_{yy}\frac{1}{\bar{r}}, \tag{105}$$
$$\frac{J_\varphi}{l_z} = \frac{\mathcal{F}^{\text{PK}}_\varphi}{l_z} = -2K^e_{yy}\frac{\sin 2\varphi}{\bar{r}}. \tag{106}$$

4.3.3. $L_3$-integral

The only non-vanishing component of the $\boldsymbol{L}$-integral (35) reads for parallel edge dislocations with parallel Burgers vectors in $y$-direction
$$L_3 = L_z = \bar{x}J_y - \bar{y}J_x - l_z\int_{-\infty}^{0} b_y\sigma_{xy}(\bar{x}+x', \bar{y})\,\mathrm{d}x'. \tag{107}$$

Substituting the $J_1$-integral (90), the $J_2$-integral (91) and the stress field $\sigma_{xy}$ from Eq. (89) into Eq. (107), the $L_3$-integral per unit dislocation length reads
$$\frac{L_3}{l_z} = -b_y b'_y D \sin 2\varphi - b_y b'_y D\int_{-\infty}^{0}\frac{\bar{y}((\bar{x}+x')^2 - \bar{y}^2)}{((\bar{x}+x')^2 + \bar{y}^2)^2}\,\mathrm{d}x'. \tag{108}$$

Using Eq. (84), Eq. (108) gives the $L_3$-integral of two parallel edge dislocations with Burgers vectors in $y$-direction per unit dislocation length
$$\frac{L_3}{l_z} = -\frac{\mu b_y b'_y}{4\pi(1-\nu)}\sin 2\varphi, \tag{109}$$

or in terms of the pre-logarithmic energy factor $K_{yy}^e$ (Eq. (98))

$$\frac{L_3}{l_z} = -K_{yy}^e \sin 2\varphi. \tag{110}$$

The mathematical structure and the physical meaning of Eqs. (109) and (110) is analogous to Eqs. (85) and (86). $L_3 = 0$ if $\varphi = \{0, \pi/2, \pi\}$.
If $b_y b'_y > 0$, then

$$\max \frac{L_3}{l_z}\left(\varphi = \frac{3\pi}{4}\right) = K_{yy}^e, \quad \min \frac{L_3}{l_z}\left(\varphi = \frac{\pi}{4}\right) = -K_{yy}^e. \tag{111}$$

If $b_y b'_y < 0$, then

$$\max \frac{L_3}{l_z}\left(\varphi = \frac{\pi}{4}\right) = |K_{yy}^e|, \quad \min \frac{L_3}{l_z}\left(\varphi = \frac{3\pi}{4}\right) = -|K_{yy}^e|. \tag{112}$$

## 5. Fundamental relations between $J$-, $M$-, and $L_3$-integrals

Here, we prove that the $J$-, $M$-, and $L_3$-integrals of straight dislocations are not independent from each other, rather than they are inherently connected to each other and we give the corresponding relations between them.

From the $J_r$-, $J_\varphi$-, $M$-, and $L_3$-integrals of screw dislocations, Eqs. (47), (48), (56), and (62); of edge dislocations with Burgers vectors in $x$-direction, Eqs. (67), (68), (76), and (85); and of edge dislocations with Burgers vectors in $y$-direction, Eqs. (92), (93), (101), and (109), respectively, we find the following *fundamental relations between the $J$-, $M$-, and $L_3$-integrals of straight dislocations in cylindrical coordinates* valid in incompatible isotropic elasticity

$$J_r = -2\frac{\partial M}{\partial \bar{r}}, \tag{113}$$

$$J_\varphi = -\frac{2}{\bar{r}}\frac{\partial M}{\partial \varphi}, \tag{114}$$

$$L_3 = -\frac{\partial M}{\partial \varphi}. \tag{115}$$

Moreover, combining Eqs. (114) and (115), we obtain a simple relation connecting directly the $J_\varphi$- and $L_3$-integrals

$$J_\varphi = \frac{2}{\bar{r}} L_3 \quad \text{or} \quad L_3 = \frac{\bar{r}}{2} J_\varphi. \tag{116}$$

Furthermore, in Cartesian coordinates, from the $J_1$-, $J_2$-, and $M$-integrals of screw dislocations, Eqs. (45), (46), and (56); of edge dislocations with Burgers vectors in $x$-direction, Eqs. (65), (66), and (76); and of edge dislocations with Burgers vectors in $y$-direction, Eqs. (90), (91), and (101), respectively, we find the following *fundamental relations between the $J$-, and $M$-integrals of straight dislocations in Cartesian coordinates*

$$J_1 = -2\frac{\partial M}{\partial \bar{x}}, \tag{117}$$

$$J_2 = -2\frac{\partial M}{\partial \bar{y}}. \tag{118}$$

Eqs. (117) and (118) can be written in the following compact form

$$J_k = -2\frac{\partial M}{\partial \bar{x}_k}, \quad k = 1, 2. \tag{119}$$

Therefore,
$$\boldsymbol{J} = -2\,\mathrm{grad}\,M\,. \tag{120}$$

We have shown above that the $J_r$-integral is twice the negative of the partial derivative of the $M$-integral with respect to the distance $\bar{r}$, the $J_\varphi$-integral is twice the negative of the partial derivative of the $M$-integral with respect to the rotation angle $\varphi$ divided by the distance $\bar{r}$, the $L_3$-integral is the negative of the partial derivative of the $M$-integral with respect to the rotation angle $\varphi$, the $J_1$-integral is twice the negative of the partial derivative of the $M$-integral with respect to the position $\bar{x}$, and the $J_2$-integral is twice the negative of the partial derivative of the $M$-integral with respect to the position $\bar{y}$. From that point of view, as far as the $\boldsymbol{J}$-, $\boldsymbol{L}$-, and $M$-integrals are concerned the $M$-integral is the primary quantity.

Moreover, it is obvious that the $\boldsymbol{J}$-, $M$-, and $L_3$-integrals are not independent from each other, quite contrary, they are inherently related by simple relations; namely the components of the $\boldsymbol{J}$-, and $\boldsymbol{L}$-integrals can be obtained by a simple differentiation of the $M$-integral with respect to the distance $\bar{r}$ or to the rotation angle $\varphi$. Therefore, if the $M$-integral is given, then the components $J_r$, $J_\varphi$, $L_3$, $J_1$, and $J_2$ can be easily derived from the relations (113), (114), (115), (117), and (118), respectively.

It should be mentioned that such fundamental relations between the $\boldsymbol{J}$-, $\boldsymbol{L}$-, and $M$-integrals have not been given before in the literature of dislocations. A single work where an analogous relation only to Eq. (115) is derived, is the work by Chen (2001a, 2002) giving the $\boldsymbol{L}$-integral as the negative of the half of the partial derivative of the $M$-integral with respect to the rotation angle for a central crack of length $2a$ in an infinite elastic body under remote uniform loading. One should keep in mind that the geometry of the different defects as well as the fact that the present study of dislocations is in the framework of incompatible elasticity taking into account the plastic fields whereas the study of cracks is in the framework of compatible elasticity, influence the explicit form of the results.

### 6. Relations between $\boldsymbol{J}$-, $M$-, $L_3$-integrals and the interaction energy

In this subsection, we connect the $\boldsymbol{J}$-, $M$-, and $L_3$-integrals of straight dislocations with the interaction energy proving among others that the interaction energy plays the physical role of a "dislocation potential" for the Peach-Koehler force. At the end of this subsection, connections and comparisons of the obtained relations with similar relations in fracture mechanics are given and discussed.

We have proven that for parallel screw dislocations and parallel edge dislocations with Burgers vectors in $x$-direction or in $y$-direction, (that is for straight dislocations with parallel dislocation lines and parallel Burgers vectors), the configurational work $\mathcal{W}^{\mathrm{PK}}$ is constant (see Eqs. (54), (74), and (99)). Recalling the fact that $M = \mathcal{W}^{\mathrm{PK}} + U_{\mathrm{d}}$ (see Eq. (19)) and using Eq. (22), we obtain

$$\frac{\partial M}{\partial \bar{x}_k} = \frac{\partial U_{\mathrm{d}}}{\partial \bar{x}_k} = \frac{1}{2}\frac{\partial U_{\mathrm{int}}}{\partial \bar{x}_k}\,. \tag{121}$$

Using Eq. (121), Eq. (119) gives that

$$J_k = -\frac{\partial U_{\mathrm{int}}}{\partial \bar{x}_k}\,, \quad k=1,2\,. \tag{122}$$

Since the $\boldsymbol{J}$-integral or equivalently the Peach-Koehler force is the negative gradient of $U_{\mathrm{int}}$

$$\boldsymbol{J} \equiv \boldsymbol{\mathcal{F}}^{\mathrm{PK}} = -\mathrm{grad}\,U_{\mathrm{int}}\,, \tag{123}$$

one can say that the interaction energy plays the physical role of a "dislocation potential" for the $\boldsymbol{J}$-integral or equivalently the Peach-Koehler force. As a consequence, we have that[3]

$$(\mathrm{curl}\,\boldsymbol{J})_z \equiv (\mathrm{curl}\,\boldsymbol{\mathcal{F}}^{\mathrm{PK}})_z = 0\,. \tag{124}$$

---

[3] $(\mathrm{curl}\,\boldsymbol{J})_z = \epsilon_{zij}\partial_i J_j$.

Thus, the $\boldsymbol{J}$-integral (122) is irrotational. Moreover, if we calculate the divergence of the $\boldsymbol{J}$-integral from Eq. (123), we conclude that the *dislocation potential $U_{int}$ satisfies the following two-dimensional Poisson equation*

$$\Delta U_{\text{int}} = -\operatorname{div} \boldsymbol{J} \equiv -\operatorname{div} \boldsymbol{\mathcal{F}}^{\text{PK}}, \tag{125}$$

where $\Delta$ denotes the Laplacian.

Since the configurational work $\mathcal{W}^{\text{PK}}$ of straight dislocations is independent of the distance $\bar{r}$ and the angle $\varphi$, it holds

$$\frac{\partial M}{\partial \bar{r}} = \frac{\partial U_{\text{d}}}{\partial \bar{r}} = \frac{1}{2} \frac{\partial U_{\text{int}}}{\partial \bar{r}} \tag{126}$$

and

$$\frac{\partial M}{\partial \varphi} = \frac{\partial U_{\text{d}}}{\partial \varphi} = \frac{1}{2} \frac{\partial U_{\text{int}}}{\partial \varphi}. \tag{127}$$

Using Eqs. (126) and (127), the $J_r$-, $J_\varphi$-, and $L_3$-integrals (Eqs. (113)–(115)) can also be expressed in terms of the interaction energy $U_{\text{int}}$, as follows

$$J_r = -\frac{\partial U_{\text{int}}}{\partial \bar{r}}, \tag{128}$$

$$J_\varphi = -\frac{1}{\bar{r}} \frac{\partial U_{\text{int}}}{\partial \varphi}, \tag{129}$$

$$L_3 = -\frac{1}{2} \frac{\partial U_{\text{int}}}{\partial \varphi}. \tag{130}$$

The $M$-integral of straight dislocations (37) can be also written entirely in terms of the interaction energy, using Eq. (122), as follows

$$M = -\bar{x}_k \frac{\partial U_{\text{int}}}{\partial \bar{x}_k} + \frac{1}{2} U_{\text{int}}. \tag{131}$$

The relations (122), (128)–(130), and (131) lead us to the conclusion that we can derive the $\boldsymbol{J}$-, $L_3$-, and $M$-integrals directly from the interaction energy. It should be pointed out that the important fact that allows us to obtain the aforementioned relations is that the configurational work $\mathcal{W}^{\text{PK}}$ produced by the Peach-Koehler force of straight dislocations is constant.

**Remark 6.** At this point, it is interesting to compare the above relations valid for straight dislocations and the corresponding relations valid for the specific problem of a plane central crack of length $2a$ in the framework of fracture mechanics. To the best of authors' knowledge only for this specific problem, analogous relations have been derived in the literature of fracture mechanics. However, even if these two problems (three-dimensional incompatible dislocation problem and two-dimensional compatible crack problem) are different, it is worth noticing some similarities between them and to point out some differences. Let us first start by observing similarities as well as differences between the interaction energy $U_{\text{int}}$ for straight dislocations (Eq. (36)) and the so-called "crack energy" $U_{\text{c}}$ for a plane crack of length $2a$ with traction-free surfaces (Eq. (23)) as defined by Golebiewska-Herrmann and Herrmann (1981b)

$$U_{\text{c}} = \frac{1}{2} \int_{-a}^{a} \sigma_{yj}^{A} \Delta u_j(x) \, \mathrm{d}x, \tag{132}$$

where $\Delta u_j(x)$ is the discontinuity in the displacement across the crack and $\sigma_{yj}^{A}$ is an applied plane homogeneous static stress field. For the considered problem, it holds (Golebiewska-Herrmann and Herrmann,

1981a; Pak *et al.*, 1983)

$$J_1 \equiv J = -\frac{\partial U_\mathrm{c}}{\partial a}\,, \tag{133}$$

$$L = -\frac{\partial U_\mathrm{c}}{\partial \phi}\,, \tag{134}$$

$$M = -2a\,\frac{\partial U_\mathrm{c}}{\partial a}\,, \tag{135}$$

where $\phi$ is in this case a small angle with respect to the $x$-axis.

Now, it is easy to see that the relation (122) for $k=1$ and the relation (133), giving the $J_1$-integral as the negative derivative of the corresponding energy ($U_\mathrm{int}$ or $U_\mathrm{c}$) with respect to the corresponding position, are analogous. Next, the relations (130) and (134), that give the $L$-integral as the negative derivative of the corresponding energy with respect to the corresponding angle, are also analogous except for the factor $1/2$. Concerning the relations (131) and (135) that relate the $M$-integral[4] with the corresponding energy, one can justify the differences recalling the fact that the $M$-integral (131) holds for a three-dimensional dislocation problem (incompatible elasticity), whereas the $M$-integral (135) holds for a two-dimensional crack problem (compatible elasticity). The $M$-integral in three-dimensions is different than the $M$-integral in two-dimensions as it has been already explained in the subsection 3.2.

## 7. Energy-release of straight dislocations

The creation of a straight dislocation with Burgers vector $b_i$ presenting discontinuity of the fields (displacement vector and plastic distortion tensor) along the $x$-direction from $-\infty$ to $0$ into an existing stress field $\sigma_{iy}$ induces a change of the total potential energy in the body, which is given by the interaction energy $U_\mathrm{int}$.

In this sense, we can define the *translational energy-release*, which is to be interpreted as the change of the total potential energy (interaction energy) under an infinitesimal translation and is denoted by $\mathcal{G}^T_k$

$$\mathcal{G}^T_k := -\frac{\partial U_\mathrm{int}}{\partial \bar{x}_k}\,, \quad k=1,2\,, \tag{136}$$

where $U_\mathrm{int}$ is given by Eq. (36).

Using Eq. (122), we conclude that

$$\mathcal{G}^T_k = J_k\,, \quad k=1,2\,, \tag{137}$$

the translational energy-release $\mathcal{G}^T_k$ is nothing but the $J_k$-integral. The minus sign in Eq. (136) has the physical interpretation of the dissipation of energy, in order to be in agreement with the more general physical interpretation of the energy-release rate in dynamics of defects as the dissipation rate (or the rate of the total energy change) due to the motion of defects or for example the creation of new crack surfaces in fracture mechanics. The translational energy-release rate for moving dislocations has been given by Clifton and Markenscoff (1981).

**Remark 7.** The relation (137) agrees with the "potential energy-release rate associated with unit crack tip advancement" for a plane central crack of length $2a$ in statics given by Pak *et al.* (1983).

Next, we proceed to define the *rotational energy-release*, which is to be interpreted as the change of the total potential energy (interaction energy) under an infinitesimal rotation (about the $z$-axis) and is denoted by $\mathcal{G}^R$

$$\mathcal{G}^R := -\frac{\partial U_\mathrm{int}}{\partial \varphi}\,. \tag{138}$$

---

[4]In Golebiewska-Herrmann and Herrmann (1981a) the factor 2 is missing; a misprint that can be checked through the corresponding relations in Pak *et al.* (1983).

Combining Eqs. (138) and (130), we obtain that the rotational energy-release $\mathcal{G}^R$ equals twice the value of the $L_3$-integral

$$\mathcal{G}^R = 2L_3. \tag{139}$$

**Remark 8.** We mention that an analogous result to Eq. (139) is also true for a plane central crack of length $2a$ with traction-free surfaces in statics, where the rotational energy-release is identical to the $L$-integral except for a change in sign (Golebiewska-Herrmann and Herrmann, 1981b).

Therefore, we arrive to the conclusion that:

- The *translational energy-release* $\mathcal{G}_k^T$ *of straight dislocations* is identical to the $J_k$-integral.
- The *rotational energy-release* $\mathcal{G}^R$ *of straight dislocations* equals twice the value of the $L_3$-integral.

## 8. $J_k$-stability criterion

In this section, we are interested in examining the stability of straight edge dislocations with respect to glide. The equilibrium of dislocations is considered as stable when the interaction energy $U_{\text{int}}$ at the force equilibrium positions possesses a minimum, and unstable when it possesses a maximum. Since the movements glide and climbing are independent from each other due to the crystal structure, and climbing can be realized only at very high temperatures and only in interaction with point defects; only glide has been studied here. The glide direction is given by the direction of the Burgers vector. For finding the minima/maxima of the interaction energy we have:

i) The interaction energy $U_{\text{int}}$ possesses a minimum at $(\bar{x}_0, \bar{y}_0)$, if

$$\frac{\partial U_{\text{int}}}{\partial \bar{x}_k} = 0 \quad \text{and} \quad \frac{\partial^2 U_{\text{int}}}{\partial \bar{x}_k^2} > 0 \quad \text{at } (\bar{x}_0, \bar{y}_0), \tag{140}$$

where $k$ is fixed ($k = 1$ or $k = 2$)[5].

ii) The interaction energy $U_{\text{int}}$ possesses a maximum at $(\bar{x}_0, \bar{y}_0)$, if

$$\frac{\partial U_{\text{int}}}{\partial \bar{x}_k} = 0 \quad \text{and} \quad \frac{\partial^2 U_{\text{int}}}{\partial \bar{x}_k^2} < 0 \quad \text{at } (\bar{x}_0, \bar{y}_0), \tag{141}$$

where $k$ is fixed ($k = 1$ or $k = 2$).

Using the relation (122) between the $J_k$-integral and the interaction energy $U_{\text{int}}$, the above conditions for minima/maxima can be rewritten in terms of the $J_k$-integral as follows:

i) The interaction energy $U_{\text{int}}$ possesses a minimum at $(\bar{x}_0, \bar{y}_0)$, if

$$J_k = 0 \quad \text{and} \quad \frac{\partial J_k}{\partial \bar{x}_k} < 0 \quad \text{at } (\bar{x}_0, \bar{y}_0), \tag{142}$$

where $k$ is fixed ($k = 1$ or $k = 2$).

ii) The interaction energy $U_{\text{int}}$ possesses a maximum at $(\bar{x}_0, \bar{y}_0)$, if

$$J_k = 0 \quad \text{and} \quad \frac{\partial J_k}{\partial \bar{x}_k} > 0 \quad \text{at } (\bar{x}_0, \bar{y}_0), \tag{143}$$

where $k$ is fixed ($k = 1$ or $k = 2$).

---

[5]Note that the $k$ is fixed because the derivatives have to be taken only with respect to the glide direction.

The above propositions can be formulated in a *stability criterion in terms of the $J_k$-integral*, which we shortly call *$J_k$-stability criterion for straight dislocations*:

At the equilibrium positions $(\bar{x}_0, \bar{y}_0)$ which are given by the *force equilibrium condition* of vanishing $J_k$-integral or vanishing Peach-Koehler force $\mathcal{F}_k^{\text{PK}}$

$$J_k = 0, \quad k = 1 \text{ or } k = 2, \tag{144}$$

the equilibrium (at these points) can be characterized by the following *stability condition*[6]

$$\frac{\partial J_k}{\partial \bar{x}_k}(\bar{x}_0, \bar{y}_0) \begin{cases} > 0 & : \text{ unstable} \\ < 0 & : \text{ stable.} \end{cases} \tag{145}$$

The first advantage of the $J_k$-stability criterion (Eqs. (144) and (145)) in comparison with the original energy-based stability criterion (Eq. (140) or Eq. (141)) is that one needs only the stress fields itself for the calculation of the $J_k$-integral and no further integration of the stress fields is needed as for the calculation of the interaction energy $U_{\text{int}}$. Second, the $J_k$-stability criterion includes only first order derivatives instead of second order derivatives. In the following subsections, we give an application of the $J_k$-stability criterion in order to find the stable and unstable equilibrium positions of the two edge-dislocation problems considered in this work.

### 8.1. Parallel edge dislocations with Burgers vectors in x-direction

The $J_1$-integral (65), which is the glide force between two parallel edge dislocations with Burgers vector in $x$-direction, reads

$$\frac{J_1}{l_z} = \frac{2K_{xx}^e}{\bar{r}} \cos\varphi \cos 2\varphi. \tag{146}$$

Substituting Eq. (146) into the condition (144), we find that the glide force is zero at three positions, $\varphi = \{\pi/4, \pi/2, 3\pi/4\}$, which are the equilibrium positions (see Fig. 11).

The stability condition (145) reads

$$\frac{1}{l_z}\frac{\partial J_1}{\partial \bar{x}} = -\frac{2K_{xx}^e}{\bar{r}^2}\left(\cos^4\varphi - 6\cos^2\varphi \sin^2\varphi + \sin^4\varphi\right). \tag{147}$$

The $\varphi$-dependence of Eq. (147) is plotted in Fig. 12. If $K_{xx}^e > 0$, that is $b_x b_x' > 0$, then

$$\frac{\partial J_1}{\partial \bar{x}}\left(\varphi = \frac{\pi}{4}\right) > 0, \quad \frac{\partial J_1}{\partial \bar{x}}\left(\varphi = \frac{3\pi}{4}\right) > 0, \tag{148}$$

whereas

$$\frac{\partial J_1}{\partial \bar{x}}\left(\varphi = \frac{\pi}{2}\right) < 0. \tag{149}$$

Therefore, according to the $J_k$-stability criterion, for two parallel edge dislocations with $b_x b_x' > 0$:

- The position $\varphi = \pi/2$ is the stable equilibrium position against glide.
- The positions $\varphi = \{\pi/4, 3\pi/4\}$ are the unstable equilibrium positions.

Reversely, for two parallel edge dislocations with $b_x b_x' < 0$:

- The positions $\varphi = \{\pi/4, 3\pi/4\}$ are the stable equilibrium positions against glide.

---

[6]If $\frac{\partial J_k}{\partial \bar{x}_k}(\bar{x}_0, \bar{y}_0) = 0$, then the stability of the equilibrium is undetermined (neutral equilibrium) and higher-order derivatives are needed (see, e.g., Ziegler (1998)). This case is beyond the scope of this work.

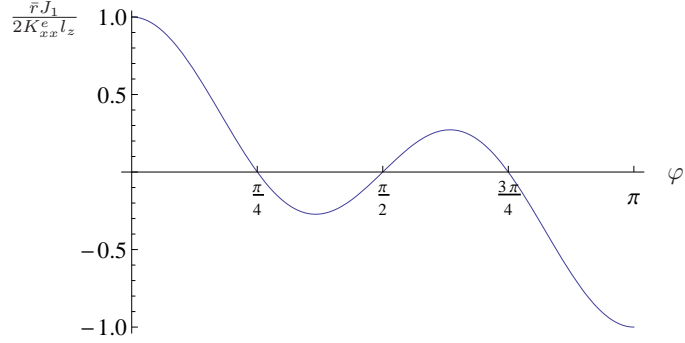

Figure 11: Normalized $J_1$-integral $\times \bar{r}$ of two parallel edge dislocations with $b_x b'_x > 0$ against $\varphi$.

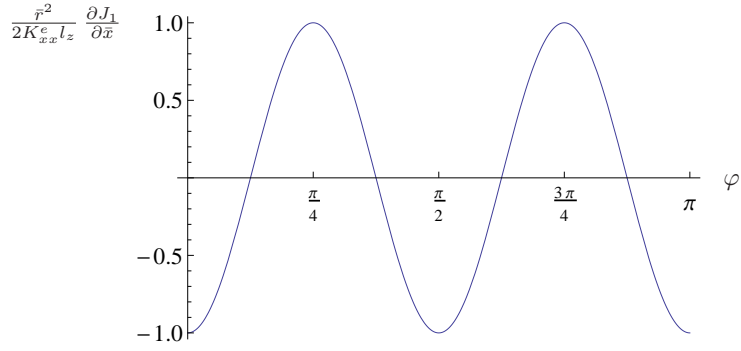

Figure 12: Normalized $\partial J_1/\partial \bar{x} \times \bar{r}^2$ of two parallel edge dislocations with $b_x b'_x > 0$ against $\varphi$.

- The position $\varphi = \pi/2$ is the unstable equilibrium position.

The two parallel edge dislocations, if free to glide, will tend to align themselves in the stable equilibrium positions. The stable and unstable equilibrium positions for parallel edge dislocations with $b_x b'_x > 0$ and $b_x b'_x < 0$ are in full agreement with those given in the textbooks of Cottrell (1953); Nabarro (1967); Kovács and Zsoldos (1973), and Lardner (1974) based on the study of the sign of the component $\mathcal{F}_x^{\text{PK}}$ of the Peach-Koehler force in the neighborhoods of the equilibrium positions.

What is worth examining furthermore is the value of the $L_3$-integral at the stable equilibrium positions. For two edge dislocations with $b_x b'_x > 0$, the $L_3$-integral is zero at the stable equilibrium position $\varphi = \pi/2$, $L_3(\varphi = \frac{\pi}{2}) = 0$. In contrast to the edge dislocations with $b_x b'_x > 0$, the $L_3$-integral is not zero at the stable equilibrium positions $\varphi = \pi/4$ and $\varphi = 3\pi/4$ of two edge dislocations with $b_x b'_x < 0$. Namely,

$$\frac{L_3}{l_z}\left(\varphi = \frac{\pi}{4}\right) = -|K_{xx}^e|, \quad \frac{L_3}{l_z}\left(\varphi = \frac{3\pi}{4}\right) = |K_{xx}^e|. \tag{150}$$

We arrive to the conclusion, that only for parallel edge dislocations with $b_x b'_x > 0$, the component of the configurational force ($J_1$: glide force) and the component of configurational vector moment $L_3$ are zero at the stable equilibrium positions, since $J_1(\varphi = \frac{\pi}{2}) = 0$ and $L_3(\varphi = \frac{\pi}{2}) = 0$.

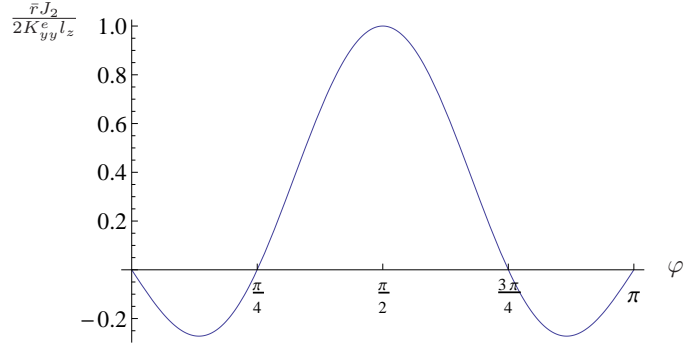

Figure 13: Normalized $J_2$-integral $\times \bar{r}$ of two parallel edge dislocations with $b_y b'_y > 0$ against $\varphi$.

*8.2. Parallel edge dislocations with Burgers vectors in y-direction*

The $J_2$-integral (91), which is the glide force between two parallel edge dislocations with Burgers vector in $y$-direction, reads

$$\frac{J_2}{l_z} = -\frac{2K^e_{yy}}{\bar{r}} \sin\varphi \cos 2\varphi\,. \tag{151}$$

Substituting Eq. (151) into the condition (144), we find that the glide force is zero at four positions, $\varphi = \{0,\ \pi/4,\ 3\pi/4,\ \pi\}$, which are the equilibrium positions (see Fig. 13).

The stability condition (145) reads now

$$\frac{1}{l_z}\frac{\partial J_2}{\partial \bar{y}} = -\frac{2K^e_{yy}}{\bar{r}^2}\left(\cos^4\varphi - 6\cos^2\varphi \sin^2\varphi + \sin^4\varphi\right). \tag{152}$$

The $\varphi$-dependence of Eq. (152) is the same as for Eq. (147) plotted in Fig. 12. If $K^e_{yy} > 0$, that is $b_y b'_y > 0$, then

$$\frac{\partial J_2}{\partial \bar{y}}\left(\varphi = \frac{\pi}{4}\right) > 0\,,\quad \frac{\partial J_2}{\partial \bar{y}}\left(\varphi = \frac{3\pi}{4}\right) > 0\,, \tag{153}$$

whereas

$$\frac{\partial J_2}{\partial \bar{y}}(\varphi = 0) < 0\,,\quad \frac{\partial J_2}{\partial \bar{y}}(\varphi = \pi) < 0\,. \tag{154}$$

Therefore, according to the $J_k$-stability criterion, for two parallel edge dislocations with $b_y b'_y > 0$:

- The positions $\varphi = \{0,\ \pi\}$ are the stable equilibrium positions against glide.
- The positions $\varphi = \{\pi/4,\ 3\pi/4\}$ are the unstable equilibrium positions.

Reversely, for two parallel edge dislocations with $b_y b'_y < 0$:

- The positions $\varphi = \{\pi/4,\ 3\pi/4\}$ are the stable equilibrium positions against glide.
- The positions $\varphi = \{0,\ \pi\}$ are the unstable equilibrium positions.

In addition, for two parallel edge dislocations with $b_y b'_y > 0$, the $L_3$-integral is zero at the stable equilibrium positions, $L_3(\varphi = 0) = 0$ and $L_3(\varphi = \pi) = 0$. Whereas for two parallel edge dislocations with $b_y b'_y < 0$, the $L_3$-integral receives its maximum and minimum values, that is

$$\frac{L_3}{l_z}\left(\varphi = \frac{\pi}{4}\right) = |K^e_{yy}|\,,\quad \frac{L_3}{l_z}\left(\varphi = \frac{3\pi}{4}\right) = -|K^e_{yy}|\,. \tag{155}$$

Therefore, again only for edge dislocations with $b_y b'_y > 0$, both the component of the configurational force ($J_2$: glide force) and the component of the configurational vector moment $L_3$ are zero at the stable equilibrium positions.

## 9. Conclusions

In this work, we have presented a methodology for the analysis of the interaction of dislocations in isotropic media using the framework of Eshelbian dislocation mechanics focusing on the utility of the $\boldsymbol{J}$-, $M$-, and $\boldsymbol{L}$-integrals. We have derived the closed-form expressions of the $\boldsymbol{J}$-, $M$-, and $L_3$-integrals for parallel screw dislocations with Burgers vectors in $z$-direction and parallel edge dislocations with Burgers vectors in $x$-direction as well as in $y$-direction. The results are gathered in Table 1. The values of the $J_1$-, $J_2$-, $M$- and $L_3$-integrals per unit dislocation length at the stable (edge) equilibrium positions with Burgers vectors in $x$-direction as well as in $y$-direction are given in Table 2 and Table 3, respectively. Furthermore, it becomes evident that the appropriate framework for studying the $\boldsymbol{J}$- , $\boldsymbol{L}$-, and $M$-integrals for dislocations is the three-dimensional, incompatible, linear elasticity. The consideration of two concepts is essential; that of the plastic fields and that of the space-dimension ($d = 3$).

The main conclusions concerning the $\boldsymbol{J}$-, $M$-, and $\boldsymbol{L}$-integrals for straight dislocations (per unit dislocation length) in isotropic materials are best underlined by the following points:

- The $\boldsymbol{J}$-integral of dislocations is the Peach-Koehler force (interaction force) between two dislocations.

- The $\boldsymbol{J}$-integral of screw dislocations depends, only on the distance between the two screw dislocations. The $\boldsymbol{J}$-integral of edge dislocations depends on the distance and on the angle between the two edge dislocations.

- The $M$-integral between two straight dislocations per unit dislocation length is the half of the corresponding interaction energy between the two dislocations per unit dislocation length plus twice the corresponding pre-logarithmic energy factor. This result gives to the $M$-integral the physical interpretation of the interaction energy between the two straight dislocations.

- The configurational work produced by the Peach-Koehler force for straight dislocations (per unit dislocation length) is constant, and equals twice the corresponding pre-logarithmic energy factor.

- The $M$-integral between two parallel screw dislocations depends only on the distance between the two dislocations. The $M$-integral between two parallel edge dislocations depends on the distance and on the angle between the two dislocations.

- The $L_3$-integral of two straight dislocations is the $z$-component of the configurational vector moment or the rotational moment about the $z$-axis caused by the interaction between the two dislocations.

- The $L_3$-integral between two screw dislocations is zero. The $L_3$-integral between two parallel edge dislocations depends only on the angle $\varphi$ between the two dislocations. It is interesting to notice that the $L_3$-integral is a function of $\sin 2\varphi$ like the $L$-integral for cracks and cavities under various loading conditions as given by Pak *et al.* (2012) (see Table 1).

- Fundamental relations between the $\boldsymbol{J}$-, $L_3$-, and $M$-integrals of straight dislocations have been found and they show that the $\boldsymbol{J}$-, $L_3$-, and $M$-integrals are not independent. If the $M$-integral is given, then the $J_1$-, $J_2$-, $J_r$-, $J_\varphi$-, and $L_3$-integrals can be easily calculated from it. From that point of view, the $M$-integral is of primary importance.

- The $\boldsymbol{J}$-, $L_3$-, and $M$-integrals of two straight dislocations are directly related with the interaction energy between the two dislocations.

- The translational energy-release $\mathcal{G}_k^T$ of straight dislocations is identical to the $J_k$-integral.

- The rotational energy-release $\mathcal{G}^R$ of straight dislocations equals twice the value of the $L_3$-integral.

- Only for parallel edge dislocations with $b_x b'_x > 0$ and $b_y b'_y > 0$, the component of the configurational force ($J_1 = \mathcal{F}_x^{\text{PK}}$ in the case $b_x b'_x > 0$, $J_2 = \mathcal{F}_y^{\text{PK}}$ in the case $b_y b'_y > 0$) and the component of the configurational vector moment $L_3$ are zero at the stable (glide) equilibrium positions (see Tables 2 and 3).

Table 1: Closed-form expressions of $\boldsymbol{J}$-, $M$-, and $L_3$-integrals for parallel straight screw and edge dislocations.

| type | $\dfrac{J_1}{l_z}$ | $\dfrac{J_2}{l_z}$ | $\dfrac{M}{l_z}$ | $\dfrac{L_3}{l_z}$ |
|---|---|---|---|---|
| $(b_z, b'_z)$ | $2K^s_{zz}\dfrac{\cos\varphi}{\bar r}$ | $2K^s_{zz}\dfrac{\sin\varphi}{\bar r}$ | $K^s_{zz}\left[2-\ln\dfrac{\bar r}{L}\right]$ | $0$ |
| $(b_x, b'_x)$ | $2K^e_{xx}\dfrac{\cos\varphi\cos 2\varphi}{\bar r}$ | $2K^e_{xx}\dfrac{\sin\varphi(2+\cos 2\varphi)}{\bar r}$ | $K^e_{xx}\left[2-\ln\dfrac{\bar r}{L}-\sin^2\varphi\right]$ | $K^e_{xx}\sin 2\varphi$ |
| $(b_y, b'_y)$ | $2K^e_{yy}\dfrac{\cos\varphi(2-\cos 2\varphi)}{\bar r}$ | $-2K^e_{yy}\dfrac{\sin\varphi\cos 2\varphi}{\bar r}$ | $K^e_{yy}\left[2-\ln\dfrac{\bar r}{L}+\sin^2\varphi\right]$ | $-K^e_{yy}\sin 2\varphi$ |

Table 2: $\boldsymbol{J}$-, $M$-, and $L_3$-integrals at the stable (glide) equilibrium positions for parallel edge dislocations with Burgers vectors in $x$-direction.

| $b_x b'_x$ | $\varphi$ | $\dfrac{J_1}{l_z}$ | $\dfrac{J_2}{l_z}$ | $\dfrac{M}{l_z}$ | $\dfrac{L_3}{l_z}$ |
|---|---|---|---|---|---|
| $>0$ | $\dfrac{\pi}{2}$ | $0$ | $2K^e_{xx}\dfrac{1}{\bar r}$ | $K^e_{xx}\left[2-\ln\dfrac{\bar r}{L}-1\right]$ | $0$ |
| $<0$ | $\dfrac{\pi}{4}$ | $0$ | $-2\sqrt{2}\,|K^e_{xx}|\dfrac{1}{\bar r}$ | $-|K^e_{xx}|\left[2-\ln\dfrac{\bar r}{L}-\dfrac{1}{2}\right]$ | $-|K^e_{xx}|$ |
| $<0$ | $\dfrac{3\pi}{4}$ | $0$ | $-2\sqrt{2}\,|K^e_{xx}|\dfrac{1}{\bar r}$ | $-|K^e_{xx}|\left[2-\ln\dfrac{\bar r}{L}-\dfrac{1}{2}\right]$ | $|K^e_{xx}|$ |

- For parallel edge dislocations with $b_x b'_x < 0$ and $b_y b'_y < 0$, the component of the configurational vector moment $L_3$ takes its maximum or minimum values at the stable (glide) equilibrium positions (see Tables 2 and 3).

We have given the unification of incompatible elasticity theory of dislocations and Eshelbian mechanics. We derived equations for general dislocation field theory from the perspective of micromechanics. We enriched the dislocation theory based on incompatible elasticity with the $\boldsymbol{J}$-, $M$-, and $\boldsymbol{L}$-integrals deriving naturally in this way important quantities for dislocation interactions, namely the Peach-Koehler force, the interaction energy and the rotational moment (torque) between dislocations. Such quantities are of great importance for a realistic simulation of discrete dislocation dynamics. Moreover, we may attach to the $\boldsymbol{J}$-, $M$-, and $\boldsymbol{L}$-integrals of dislocations a more general terminology in the framework of Eshelbian dislocation mechanics, that of "driving force", "driving work", and "driving moment", respectively. All these quantities play a significant role in dislocation-based plasticity and dislocation-based fracture mechanics. The given expressions for the "driving force", "driving work", and "driving moment" of dislocations ($\boldsymbol{J}$-, $M$-, and $\boldsymbol{L}$-integrals) can be used for finite element implementation of dislocation-mechanical problems. The presented dislocation formulation is an important contribution to dislocation theory and Eshelbian mechanics with promising application possibilities to discrete dislocation dynamics, dislocation-based fracture mechanics and other meso-scale crystal plasticity theories and computations.

Table 3: $J$-, $M$-, and $L_3$-integrals at the stable (glide) equilibrium positions for parallel edge dislocations with Burgers vectors in $y$-direction.

| $b_y b'_y$ | $\varphi$ | $\dfrac{J_1}{l_z}$ | $\dfrac{J_2}{l_z}$ | $\dfrac{M}{l_z}$ | $\dfrac{L_3}{l_z}$ |
|---|---|---|---|---|---|
| $>0$ | $0$ | $2K_{yy}^e \dfrac{1}{\bar{r}}$ | $0$ | $K_{yy}^e\bigl[2-\ln\tfrac{\bar{r}}{L}\bigr]$ | $0$ |
| $>0$ | $\pi$ | $-2K_{yy}^e \dfrac{1}{\bar{r}}$ | $0$ | $K_{yy}^e\bigl[2-\ln\tfrac{\bar{r}}{L}\bigr]$ | $0$ |
| $<0$ | $\dfrac{\pi}{4}$ | $-2\sqrt{2}\,|K_{yy}^e|\dfrac{1}{\bar{r}}$ | $0$ | $-|K_{yy}^e|\bigl[2-\ln\tfrac{\bar{r}}{L}+\tfrac{1}{2}\bigr]$ | $|K_{yy}^e|$ |
| $<0$ | $\dfrac{3\pi}{4}$ | $2\sqrt{2}\,|K_{yy}^e|\dfrac{1}{\bar{r}}$ | $0$ | $-|K_{yy}^e|\bigl[2-\ln\tfrac{\bar{r}}{L}+\tfrac{1}{2}\bigr]$ | $-|K_{yy}^e|$ |

## Acknowledgements

Eleni Agiasofitou is grateful to Georgios E. Stavroulakis for useful remarks and interesting discussions. The authors gratefully acknowledge grants from the Deutsche Forschungsgemeinschaft (Grant Nos. La1974/2-2, La1974/3-1, La1974/3-2).